\documentclass[12pt]{article}
\usepackage{amsmath}
\usepackage{graphicx,psfrag,epsf}
\usepackage{subfig}
\usepackage{enumerate}
\usepackage{natbib}

\usepackage{amssymb}
\usepackage[colorlinks,citecolor=blue,urlcolor=blue]{hyperref}
\usepackage{mathrsfs}
\usepackage{amsfonts,anysize,amsmath,indentfirst,geometry}
\usepackage{bm}
\usepackage{amsthm}
\usepackage{multirow}
\usepackage{chngpage}
\usepackage{array}
\usepackage{rotating}
\usepackage{setspace}
\usepackage{graphicx}
\usepackage[ruled]{algorithm2e}
\usepackage{epsfig}
\usepackage{amsbsy}
\usepackage{bbm}
\usepackage{booktabs}
\usepackage[table]{xcolor}
\newtheorem{theorem}{Theorem}
\newtheorem{proposition}{Proposition}

\newtheorem{remark}{\underline{\bf Remark}}

\def\boxit#1{\vbox{\hrule\hbox{\vrule\kern6pt\vbox{\kern6pt#1\kern6pt}\kern6pt\vrule}\hrule}}
\def\bqe{\begin{eqnarray}}
\def\eqe{\end{eqnarray}}

\newcommand{\blind}{0}

\begin{document}
\doublespacing

\def\spacingset#1{\renewcommand{\baselinestretch}%
{#1}\small\normalsize} \spacingset{1}

%%%%%%%%%%%%%%%%%%%%%%%%%%%%%%%%%%%%%%%%%%%%%%%%%%%%%%%%%%%%%%%%%%%%%%%%%%%%%%

\if0\blind
{
  \title{\bf Sparse-group Convoluted Smoothed Quantile Regression in High Dimension}
  \author{Jing Lv and Anxi Wang\\
    {\small School of Mathematics and Statistics, Southwest University, Chongqing, China}
    }
  \maketitle
} \fi

%\bigskip
\begin{abstract}
%\spacingset{1.45}
In many applications, covariates are naturally organized into groups, yet conventional group-selection methods often retain all variables within selected groups, failing to achieve within-group sparsity. To address this limitation, we propose an adaptive two-stage sparse-group selection procedure for high-dimensional quantile regression. The proposed method integrates convolution-smoothed quantile loss with $\ell_{2,0}$ and $\ell_0$ penalties to identify important groups and relevant variables within them, achieving sparsity at both levels. We establish theoretical guarantees for sure screening and exact support recovery at both stages, as well as consistent model-size selection using information criteria, with explicit bounds on failure probabilities. The proposed procedure is computationally efficient by restricting optimization to selected groups and active variables. Simulation studies and a breast cancer gene-expression analysis demonstrate its competitive performance in variable selection, estimation accuracy, and computational efficiency, while identifying biologically relevant genes and delivering accurate quantile predictions.
\end{abstract}

%\noindent%
\\

\section{Introduction}\label{sec:intro}
\spacingset{1.45}

Sparse group recovery has received considerable attention in a wide range of applications, including wavelet image analysis \citep{S1993}, signal processing \citep{JJL2017,TY24}, and gene analysis \citep{MSH2007}. In many practical problems, covariates are naturally partitioned into groups because they share common characteristics or because prior scientific knowledge suggests that variables within the same group may exert related effects. For example, image signals often exhibit grouped nonzero patterns, whereas assayed genes or proteins may be organized according to biological functions or pathways. Incorporating such grouping information into statistical modeling can substantially improve both model interpretability and predictive performance.

Over the past few decades, extensive research has addressed high-dimensional group selection using group $\ell_1$ and nonconvex penalties, including the group Lasso, group SCAD, and group MCP; see \cite{YL06,LAL2019,YN2020,TY24} for reviews and related developments. Although these methods offer attractive theoretical and numerical properties, tuning-parameter selection can be computationally demanding, particularly when constructing solution paths across different model sizes. Alternatively, $\ell_0$-regularized procedures directly penalize the number of selected variables, offering a transparent sparsity interpretation and favorable selection properties; see, for example, \cite{HJLL18,ZWZZW20,MY24,MY2025}. As a natural extension of the $\ell_0$ penalty to grouped variables, the $\ell_{2,0}$ penalty has also been used for group-sparse recovery. \cite{JJL2017} studied an $\ell_{2,0}$-regularized least-squares estimator, while \cite{ZZZW23} proposed a group-splicing algorithm that iteratively detects relevant groups and removes irrelevant ones. More recently, \cite{LG2025} combined rank-based methodology with an $\ell_{2,0}$ penalty to achieve robust group selection.

In many applications, however, group-wise sparsity alone is not sufficient. Practitioners are often interested not only in identifying important groups, but also in selecting the truly relevant variables within those groups. For example, although a biological pathway may be associated with a particular biological process, this does not imply that every gene in the pathway is relevant to that process. Conventional group-selection methods do not induce within-group sparsity: once a group is selected, all variables within that group are retained \citep{YL06,ZZZW23,LG2025}. To distinguish these two levels of sparsity, throughout this paper we use the terms \emph{group-wise sparsity} and \emph{within-group sparsity}. A procedure that simultaneously achieves both types of sparsity is referred to as a \emph{sparse-group selection} method.

Several methods have been proposed for sparse-group selection. For example, \cite{SFHT2013} introduced the sparse-group Lasso, which combines Lasso and group-Lasso penalties to select both groups and variables within groups. However, the method involves two tuning parameters whose joint selection can be computationally demanding. To alleviate this difficulty, \cite{LAL2019} proposed a gradient-free coordinate descent algorithm that automatically selects the tuning parameters of the sparse-group Lasso. \cite{CZZ2022} further established theoretical properties of the sparse-group Lasso, including upper and matching minimax lower bounds for estimation error. Related developments can also be found in \cite{ZZ2010,LNZ2015}.

Most existing sparse-group procedures have been developed for diverging-dimensional settings in which the dimension $p_n$ increases with the sample size $n$ but remains smaller than $n$, for example, $p_n=O(n^a)$ for some $a\in(0,1)$. Hence, they are not directly applicable to ultrahigh-dimensional settings with $p_n\gg n$. Moreover, many existing sparse-group methods are built upon least-squares loss and therefore target the conditional mean. Such procedures can be sensitive to outliers and heavy-tailed errors and may fail to capture heterogeneity in the effects of covariates across different parts of the conditional distribution.

Quantile regression (QR), introduced by \cite{KB1978}, provides a natural alternative because it is robust to outliers, accommodates skewed and heavy-tailed distributions, and offers a more comprehensive characterization of the conditional distribution of the response. High-dimensional QR based on $\ell_1$ regularization has been extensively studied; see, for example, \cite{BC2011,WWL2012,CLMY20,CFHL2022}. A major computational challenge, however, is that the classical quantile loss is nonsmooth: although QR can be formulated as a linear programming problem, such direct approaches become computationally expensive or memory-intensive when $n$ and $p_n$ are both large.

To overcome the nonsmoothness of the quantile loss, \cite{TWZ2022} proposed smoothing the subgradient of the check loss and integrating the resulting function to obtain a convex smooth loss. We refer to this approach as \emph{convoluted smoothed quantile regression} (CSQR).\cite{HPTZ2023} and \cite{YZZ2023} developed large-scale inference for CSQR, including multiplier bootstrap, confidence intervals, and hypothesis tests, in the $p_n<n$ and high-dimensional settings. Nevertheless, existing high-dimensional QR procedures mainly focus on either individual-variable selection or entire-group selection \citep{BC2011,WWL2012,PW2015,CLMY20,TWZ2022,YZZ2023}, and relatively little attention has been paid to simultaneously achieving group-wise and within-group sparsity. This limitation may lead to unnecessarily large fitted models when only a subset of variables in an active group is truly relevant.

Motivated by these considerations, we develop a robust two-stage sparse-group selection framework that combines CSQR with $\ell_{2,0}$ and $\ell_0$ penalties. The proposed procedure first identifies important groups and then performs variable selection within the retained groups, thereby exploiting structural information while avoiding the unnecessary inclusion of irrelevant within-group variables. The main contributions of this work are summarized as follows.

\begin{itemize}

\item
We first combine the convoluted smoothed quantile loss with an $\ell_{2,0}$ penalty for robust group selection. As the resulting problem is nonconvex and discontinuous, the optimality characterization in Proposition~\ref{pro.1} motivates Algorithm~\ref{alg:1}, which constructs a solution path over group sizes: active groups are updated using gradient and Hessian information and the coefficients are then re-estimated by a restricted CSQR refit, providing effective and computationally efficient dimension reduction. Under mild regularity conditions, this path enjoys  group-recovery guarantees, and the proposed GBIC consistently selects the number of active groups.

\item
We then impose an $\ell_0$-regularized CSQR on the variables in the retained groups. The coordinate-wise characterization in Proposition~\ref{pro.2} motivates Algorithm~\ref{alg:2}, which identifies important variables while preserving the robustness. Restricting the search to the groups retained after screening makes this stage substantially more efficient than direct variable selection over the full predictor set. The resulting within-group path likewise admits sure screening and exact variable-recovery guarantees, and the proposed BIC consistently selects the within-group model size.

\item
We combine the two stage-specific procedures with GBIC and BIC model-size
selection to develop the fully adaptive Algorithm~\ref{alg:3}, which produces a
doubly sparse model by simultaneously identifying important groups and relevant
variables within the selected groups. This distinguishes our method from
existing high-dimensional quantile regression procedures that typically
consider only one level of sparsity
\citep{WWL2012,PW2015,CLMY20,TWZ2022,YZZ2023}, as well as sparse-group methods
mainly based on least-squares loss \citep{ZZ2010,SFHT2013,LAL2019}.
The two-stage active-set strategy also restricts optimization to progressively
reduced parameter spaces, avoiding an exhaustive search over all $p_n$
variables and substantially reducing the computational burden.
\end{itemize}

%The rest of the paper is organized as follows. In Section~\ref{sect.2}, we introduce the group support detection and root-finding procedure in Algorithm~\ref{alg:1} for achieving group-wise sparsity, and then develop Algorithm~\ref{alg:2} for selecting important variables within the identified active groups. We further combine these two procedures with GBIC and BIC to construct the adaptive sparse-group selection method in Algorithm~\ref{alg:3}. Section~\ref{sect.3A} presents implementation details and simulation studies evaluating the finite-sample performance of the proposed methods. Section~\ref{sect.4A} analyzes a breast cancer gene-expression dataset to demonstrate the practical utility of the proposed framework. Section~\ref{sect.5A} concludes the paper with additional discussion. Proofs of the theoretical results are provided in the Supplementary Material.

\section{Methodology}\label{sect.2}
\subsection{Model Set-up and Notation}\label{sect.2.1}

We consider a univariate response $\mathrm{y}\in \mathbb{R}$ and a $p_n$-dimensional covariate vector $\bm {\mathrm{x}} =\left(\mathrm{x}_{1},\cdots,\mathrm{x}_{p_n}\right)^{\mathrm{T}}$.
Assume that the $p_n$ predictors can be naturally divided into $J$ non-overlapping groups $\mathcal{G}_1,\cdots,\mathcal{G}_J$ such that $\bigcup\nolimits_{j = 1}^J \mathcal{G}_j =\left\{1,\cdots,p_n\right\}\triangleq \Omega$ and $\mathcal{G}_j\bigcap\mathcal{G}_{j'}=\varnothing$ for $j\neq j'$.
The size of the $j$th group is defined by $p_j$ (i.e., $|\mathcal{G}_{j}|=p_j$) for $j\in \mathcal{C}$, and hence $\sum_{j=1}^J p_j=p_n$, where $\mathcal{C}= \left\{1,\ldots,J\right\}$.
Consider the classical multi-linear model with $J$ non-overlapping groups
\begin{align}\label{eq1}
\mathrm{y}=\alpha+\sum_{j =1}^{J}{\bm {\mathrm{x}}_{\mathcal{G}_j}^{\mathrm{T}}\bm\beta_{\mathcal{G}_j}}+\epsilon,
\end{align}
where
$\bm {\mathrm{x}}_{\mathcal{G}_j}=\big( {\mathrm{x}}_{k} , k \in \mathcal{G}_j \big)^{\mathrm{T}} \in\mathbb{R}^{ {p_j}}$ and $\bm\beta_{\mathcal{G}_j}=\big(\beta_{j'},j' \in \mathcal{G}_j\big)^{\mathrm{T}} \in \mathbb{R}^{p_j}$ are the covariate vector and the regression coefficient vector respectively for the $j$th group, $\alpha$ is an intercept and {$\epsilon$} is the random error satisfying $\mathbb{P}\{\epsilon\leq 0|\bm {\mathrm{x}}\}=\tau$.
(otherwise, we can shift the intercept to be $\alpha-F_{\epsilon|\bm {\mathrm{x}}}^{-1}(\tau)$ so that this assumption still holds, where $F_{\epsilon|\bm {\mathrm{x}}}(\cdot)$ is the cumulative distribution function of $\epsilon$ given $\bm {\mathrm{x}}$).
Under the above assumption, the $\tau$th conditional quantile of $\mathrm{y}$ given $\bm {\mathrm{x}}$ is given by $Q_{\tau}\left(y|\bm {\mathrm{x}}\right)=\alpha+\sum_{j =1}^{J}{\bm {\mathrm{x}}_{\mathcal{G}_j}^{\mathrm{T}}\bm\beta_{\mathcal{G}_j}}$. Throughout the paper, we focus on the high-dimensional scenario in which the number of features $p_n$ is substantially greater than the sample size $n$.

We first introduce some useful notation for the content in the following.
For any $\mathcal{S}\subset \mathcal{C}$, define $\mathcal{G}_{\mathcal{S}}=\bigcup_{j\in \mathcal{S}}\mathcal{G}_j$, $|\mathcal{S}|^\diamond=\sum_{j \in\mathcal{S}}p_j$, $\bm\beta_{\mathcal{S}}=\big(\bm\beta_{\mathcal{G}_j}^{\mathrm{T}},j \in \mathcal{S}\big)^{\mathrm{T}} \in \mathbb{R}^{| \mathcal{S}|^\diamond}$ with $\bm\beta_{\mathcal{G}_j}=\big(\beta_{j'},j' \in \mathcal{G}_j\big)^{\mathrm{T}} \in \mathbb{R}^{p_j}$ and $\bm {\mathrm{x}}_{\mathcal{S}}=\big(\bm {\mathrm{x}}_{\mathcal{G}_j}^{\mathrm{T}}, j \in \mathcal{S} \big)^{\mathrm{T}}\in \mathbb{R}^{| \mathcal{S}|^\diamond}$ with $\bm {\mathrm{x}}_{\mathcal{G}_j}=\big({\mathrm{x}}_{k},k\in \mathcal{G}_j \big)^{\mathrm{T}} \in \mathbb{R}^{p_j}$, where ${\mathrm{x}}_{k}$ is the $k$th component of $\bm {\mathrm{x}}$. Assume that the true regression coefficient $\bm{\mathrm{\beta}}^*$
possesses a double sparse structure between groups and within groups. Specifically, let $\mathcal{S}^* = \mathrm{gsupp}(\bm{\mathrm{\beta}}^*) = \{ j\in \mathcal{C}: \|\bm{\mathrm{\beta}}^*_{\mathcal{G}_j}\|_2 \neq 0\}$ be the set of active groups and $\mathcal{A}^* = \mathrm{supp}(\bm{\mathrm{\beta}}^*) = \{ k \in \mathcal{G}_{\mathcal{S}^*} : \beta^*_k \neq 0\}$ be the set of nonzero variables in active groups,
where $\mathcal{G}_{\mathcal{S}^*} = \bigcup_{j\in \mathcal{S}^*}\mathcal{G}_j$ and $\|\cdot\|_2$ is the $\ell_{2}$ norm of a vector. Further, we assume $|\mathcal{S}^*|=s^*$ and $|\mathcal{A}^*|=r^*$.
For any two sets $\mathcal{A}$ and $\mathcal{B}$, let $\mathcal{A}\backslash \mathcal{B}=\mathcal{A}\cap \mathcal{B}^c=\mathcal{A}-\mathcal{A}\cap \mathcal{B}$ and denote $|\mathcal{A}|$ as
its size of $\mathcal{A}$, where $\mathcal{B}^c$ is the complement of $\mathcal{B}$.
Define $\mathcal{F}^* = \mathcal{C}\backslash \mathcal{\mathcal{S}^*}$ and $\mathcal{I}^* =\mathcal{G}_{\mathcal{S}^*}\backslash\mathcal{A}^* $. Under the above assumption, we have $|\mathcal{F}^*|=J-s^*$, $|\mathcal{I}^*|=|\mathcal{S}^*|^\diamond-r^*=\sum_{j \in\mathcal{S}^*}p_j-r^*$, $\mathcal{C}=\mathcal{S}^*\bigcup\mathcal{F}^*$ and $\mathcal{G}_{\mathcal{S}^*}=\mathcal{A}^*\bigcup\mathcal{I}^*$, and hence $\bm{\mathrm{\beta}}^*$ can be written as $\bm{\mathrm{\beta}}^* = (\bm{\mathrm{\beta}}_{\mathcal{S}^*}^{*\mathrm{T}},\, \bm{\mathrm{\beta}}_{\mathcal{F}^*}^{*\mathrm{T}})^\mathrm{T}$, where $\bm{\mathrm{\beta}}_{\mathcal{S}^*}^* =(\bm{\mathrm{\beta}}_{\mathcal{A}^*}^{*\mathrm{T}},\, \bm{\mathrm{\beta}}_{\mathcal{I}^*}^{*\mathrm{T}})^\mathrm{T}$, $\bm{\mathrm{\beta}}_{\mathcal{F}^*}^* =\bm{0}$, $\bm\beta_{\mathcal{A}^*}^*=\big(\beta_k^*,k \in \mathcal{A}^*\big)^{\mathrm{T}} \in \mathbb{R}^{| \mathcal{A}^*|}$ and $\bm\beta_{\mathcal{I}^*}^*=\bm 0$.

\subsection{$\ell_{2,0}$-regularized CSQR for Group-wise sparsity}

Suppose that we observe $n$ independent samples
$\{(\bm{\mathrm{x}}_i,\mathrm{y}_i)\}_{i=1}^n$ from model~\eqref{eq1}. The
standard quantile regression estimator of $\alpha$ and $\bm{\mathrm{\beta}}$
is obtained from
\begin{align}\label{eq2}
\min_{\alpha,\bm{\mathrm{\beta}}}\mathcal{L}_n(\alpha,\bm{\mathrm{\beta}}),
\end{align}
where
$\mathcal{L}_n(\alpha,\bm{\mathrm{\beta}})
=
\frac{1}{n}\sum_{i=1}^n
\rho_\tau\!\left(\mathrm{y}_i-\alpha-\bm{\mathrm{x}}_i^{\mathrm T}
\bm{\mathrm{\beta}}\right)$
and $\rho_\tau(u)=u\{\tau-\mathbb{I}(u<0)\}$ is the check loss
\citep{KB1978}. Its non-smoothness, however, complicates both theoretical
analysis and gradient-based optimization.

To address this issue while preserving convexity, \cite{TWZ2022} proposed the
convoluted smoothed quantile loss
\[
\widetilde{\mathcal{L}}_{n,h}(\alpha,\bm{\mathrm{\beta}})
=
\frac{1}{n}\sum_{i=1}^n
\ell_{h}\!\left(
\mathrm{y}_i-\alpha-
\sum_{j=1}^{J}
\bm{\mathrm{x}}_{i\mathcal{G}_j}^{\mathrm T}
\bm{\mathrm{\beta}}_{\mathcal{G}_j}
\right),
\]
where
$\ell_{h}(u)
=
\int_{-\infty}^{\infty}
\rho_\tau(v)K_h(v-u)\,dv$ and
$K_h(u)=h^{-1}K(u/h)$. Here $K(\cdot)$ is a kernel function and $h>0$ is a bandwidth. The resulting
loss is twice continuously differentiable and locally strongly convex with
high probability, and can be efficiently minimized using the R package
\texttt{conquer} \citep{HPTZ2023}.

Although group Lasso, group SCAD, and group MCP can achieve group-wise
sparsity in quantile regression \citep{PW2015}, convex group penalties may
introduce substantial bias, whereas nonconvex penalties lead to more difficult
optimization. We therefore consider
\begin{align}\label{eq5A}
\min_{\alpha,\bm{\mathrm{\beta}}}
\widetilde{\mathcal{L}}_{n,h,\lambda_1}
(\alpha,\bm{\mathrm{\beta}}),
\end{align}
where $\widetilde{\mathcal{L}}_{n,h,\lambda_1}(\alpha,\bm{\mathrm{\beta}})=\widetilde{\mathcal{L}}_{n,h}(\alpha,\bm{\mathrm{\beta}})+ \lambda_1 \|\bm{\mathrm{\beta}}\|_{2,0}$, $\|\bm{\mathrm{\beta}}\|_{2,0}
= \sum_{j=1}^J  \mathbb{I} \{\|\bm{\mathrm{\beta}}_{\mathcal{G}_j}\|_2 \neq 0\}$ with $\|\cdot\|_2$ being the $\ell_{2}$ norm of a vector, and $\lambda_1>0$ controls the group-wise sparsity level.

Since the group $\ell_{2,0}$ penalty is discontinuous and
nonconvex, finding a global minimizer of \eqref{eq5A} is generally
computationally challenging. Following the block coordinate-wise
minimization approach, we say that
$(\alpha^\diamond,\bm\beta^\diamond)$ is a block
coordinate-wise minimizer of \eqref{eq5A} if
\[
\alpha^\diamondsuit
\in \mathop{\arg\min}_{\alpha}
\widetilde{\mathcal L}_{n,h}
(\alpha,\bm\beta^\diamondsuit)
\]
and, for every $j\in\mathcal C$,
\begin{align}\label{eq6}
\bm\beta_{\mathcal{G}_j}^\diamondsuit
\in
\arg\min_{\bm\beta_{\mathcal{G}_j}}
\widetilde{\mathcal{L}}_{n,h,\lambda_1}
\left(
\alpha^\diamondsuit,
\bm\beta_{\mathcal{G}_1}^\diamondsuit,\ldots,
\bm\beta_{\mathcal{G}_{j-1}}^\diamondsuit,
\bm\beta_{\mathcal{G}_j},
\bm\beta_{\mathcal{G}_{j+1}}^\diamondsuit,\ldots,
\bm\beta_{\mathcal{G}_J}^\diamondsuit
\right).
\end{align}
Further, let $\widetilde{\mathcal{L}}_{\mathcal{G}_j,h,\lambda_1}\left(\bm{\mathrm{\beta}}_{\mathcal{G}_j}\right)\triangleq
\frac{1}{n}\sum_{i=1}^{n}\ell_h(r_{i\mathcal{G}_{j}}^\diamondsuit(\bm\beta_{\mathcal{G}_j}))+\lambda_1{\left\| \bm\beta _{\mathcal{G}_j} \right\|_{2,0}}$, where $r_{i\mathcal{G}_{j}}^\diamondsuit(\bm\beta_{\mathcal{G}_j})=\mathrm{y}_{i}-\bm{\mathrm{x}}_{i\mathcal{G}_j}^{\mathrm{T}}\bm\beta_{\mathcal{G}_j}-\alpha^\diamondsuit-\sum\limits_{j'=1,j'\ne j}^J\bm{\mathrm{x}}_{i\mathcal{G}_{j'}}^{\mathrm{T}}\bm{\mathrm{\beta}}_{\mathcal{G}_{j'}}^\diamondsuit$.
Then, after removing terms independent of
$\bm\beta_{\mathcal{G}_j}$, \eqref{eq6} reduces to
\begin{align}\label{eq6A}
\bm{\mathrm{\beta}}_{\mathcal{G}_j}^\diamondsuit
\in
\arg\min_{\bm{\mathrm{\beta}}_{\mathcal{G}_j}}
\widetilde{\mathcal{L}}_{\mathcal{G}_j,h,\lambda_1}
(\bm{\mathrm{\beta}}_{\mathcal{G}_j}).
\end{align}

We introduce useful notations.
Let $\bm{\mathcal{H}}_{\mathcal{G}_j}^\diamondsuit \triangleq \nabla^2_{\mathcal{G}_j}{\widetilde{\cal L}_{n,h}}\left(\alpha^\diamondsuit, \bm{\mathrm{\beta}}^\diamondsuit \right)= \frac{{\partial ^2 {{\widetilde{\cal L}}_{n,h}}\left(\alpha, \bm\beta  \right)}}{{\partial\bm \beta_{\mathcal{G}_{j}} }{\partial \bm\beta_{\mathcal{G}_{j}}^{\mathrm{T}} }}\left| {_{\alpha=\alpha^\diamondsuit,\bm\beta  = \bm\beta^\diamondsuit}} \right.=\frac{1}{n}\sum\limits_{i=1}^{n}K_h\left(-r_{i\mathcal{G}_j}^{\diamondsuit}\left( \bm{\mathrm{\beta}}_{\mathcal{G}_j}^\diamondsuit\right) \right)\bm{\mathrm{x}}_{i\mathcal{G}_j}\bm{\mathrm{x}}_{i\mathcal{G}_j}^{\mathrm{T}}$, $\bm{d}_{\mathcal{G}_j}^\diamondsuit\triangleq -\big(\bm{\mathcal{H}}_{\mathcal{G}_j}^\diamondsuit\big)^{-1}{\nabla _{\mathcal{G}_{j}}}{\widetilde{\cal L}_{n,h}}\left(\alpha^\diamondsuit,\bm\beta^\diamondsuit\right)$,
 ${\nabla _{\mathcal{G}_{j}}}{\widetilde{\cal L}_{n,h}}\left(\alpha^\diamondsuit,\bm\beta^\diamondsuit\right)=  \frac{{\partial {{\widetilde{\cal L}}_{n,h}}\left(\alpha,\bm \beta  \right)}}{{\partial {\bm\beta _{\mathcal{G}_j}}}}\left| {_{\alpha=\alpha^\diamondsuit,\bm\beta  = \bm\beta^\diamondsuit}} \right.
=\frac{1}{n}\sum\limits_{i=1}^{n}\left( \overline{K}\left( -r_{i\mathcal{G}_j}^\diamondsuit(\bm{\mathrm{\beta}}_{\mathcal{G}_j}^\diamondsuit) /h\right) -\tau\right) \bm{\mathrm{x}}_{i\mathcal{G}_j}$ and $\overline{K}(u)=\int_{-\infty}^uK(t)dt$. Now, we derive the optimality conditions for the block coordinatewise minimizer $\bm\beta^\diamondsuit$ of problem \eqref{eq5A}, where $\bm\beta^\diamondsuit=\left(\left(\bm\beta_{\mathcal{G}_j}^\diamondsuit\right)^\mathrm{T},j\in \mathcal{C} \right)^\mathrm{T}$.

\begin{proposition}\label{pro.1}
Suppose that $(\alpha^\diamondsuit,\bm\beta^\diamondsuit)$ is a coordinate-wise minimizer of \eqref{eq5A} with respect to the intercept and each group block. Assume further that, for each $j\in\mathcal C$,
$\bm{\mathcal H}_{\mathcal G_j}^\diamondsuit$ is positive definite,
i.e., $\bm{\mathcal H}_{\mathcal G_j}^\diamondsuit\succ0$, and $\nabla^2_{\mathcal G_j}
\widetilde{\mathcal L}_{n,h}
\left(\alpha^\diamondsuit,\bm\beta_{\mathcal{G}_1}^\diamondsuit,\cdots,\bm\beta_{\mathcal{G}_{j-1}}^\diamondsuit,
	\bm\beta_{\mathcal{G}_j}, \bm\beta_{\mathcal{G}_{j+1}}^\diamondsuit,\cdots,\bm\beta_{\mathcal{G}_J}^\diamondsuit\right)
\preceq
\bm{\mathcal H}_{\mathcal G_j}^\diamondsuit$,
$\bm\beta_{\mathcal G_j}\in
[\,\bm0,\bm\beta_{\mathcal G_j}^\diamondsuit\,]
\cup
[\,\bm0,\bm d_{\mathcal G_j}^\diamondsuit\,]$.
Then $(\alpha^\diamondsuit,\bm\beta^\diamondsuit)$ satisfies
\begin{align}\label{eq7}
\left\{
\begin{array}{l}
{\mathcal S}^\diamondsuit=\{j\in\mathcal C:\|\bm\Delta_j^\diamondsuit\|_2^2\geq2\lambda_1\}\\
{\mathcal F}^\diamondsuit=\{j\in\mathcal C:\|\bm\Delta_j^\diamondsuit\|_2^2<2\lambda_1\}\\
\left(\alpha^\diamondsuit,
\bm\beta_{{\mathcal S}^\diamondsuit}^\diamondsuit\right)
\in
\mathop{\arg\min}\limits_{\alpha,
\bm\beta_{{\mathcal S}^\diamondsuit}}
\widetilde{\mathcal L}_{n,h}
\left(\alpha,\bm\beta_{{\mathcal S}^\diamondsuit}\right)\\
\bm\beta_{{\mathcal F}^\diamondsuit}^\diamondsuit=\bm0
\end{array}
\right.
\end{align}
where
$\widetilde{\mathcal L}_{n,h}
(\alpha,\bm\beta_{{\mathcal S}^\diamondsuit})
=n^{-1}\sum_{i=1}^n
\ell_h(\mathrm y_i-\alpha-
\bm x_{i{\mathcal S}^\diamondsuit}^{\rm T}
\bm\beta_{{\mathcal S}^\diamondsuit})$,
$\bm\beta_{{\mathcal S}^\diamondsuit}^\diamondsuit
=((\bm\beta_{\mathcal G_j}^\diamondsuit)^{\rm T},
j\in{\mathcal S}^\diamondsuit)^{\rm T}$,
$\bm\beta_{{\mathcal F}^\diamondsuit}^\diamondsuit
=((\bm\beta_{\mathcal G_j}^\diamondsuit)^{\rm T},
j\in{\mathcal F}^\diamondsuit)^{\rm T}$, and
$\bm\Delta_j^\diamondsuit
=(\bm{\mathcal H}_{\mathcal G_j}^\diamondsuit)^{1/2}
(\bm\beta_{\mathcal G_j}^\diamondsuit+
\bm d_{\mathcal G_j}^\diamondsuit)$.
Conversely, if $(\alpha^\diamondsuit,\bm\beta^\diamondsuit)$ satisfies \eqref{eq7}, then $(\alpha^\diamondsuit,\bm\beta^\diamondsuit)$ is a local minimizer of \eqref{eq5A}.
\end{proposition}

\begin{remark}\label{rm1_Pro1A}
Proposition \ref{pro.1} gives threshold implications for a coordinate-wise minimizer of the $\ell_{2,0}$-regularized problem \eqref{eq5A}. It partitions the group index set $\mathcal{C}$ into the active set $ {\mathcal{S}}^\diamondsuit$ and the inactive set ${\mathcal{F}}^\diamondsuit$, enforcing exact group sparsity by setting $\bm\beta_{{\mathcal{F}}^\diamondsuit}^\diamondsuit=\bm 0$ and estimating $\alpha^\diamondsuit$ and $\bm\beta_{{\mathcal{S}}^\diamondsuit}^\diamondsuit$ via unpenalized minimization of $\widetilde{\mathcal{L}}_{n,h}\left(\alpha, \bm\beta_{{\mathcal{S}}^\diamondsuit}\right)$. This result not only provides the optimal conditions of the minimization problem \eqref{eq5A} but also
further inspires the block coordinate descent algorithm to
solve \eqref{eq5A} numerically.
The proofs of Proposition \ref{pro.1} are provided in the Supplementary Material.
\end{remark}

\begin{remark}\label{rm1_Pro1B}
In practice, it is challenging to select a value of $\lambda_1$ that ensures $|\mathcal{S}| = s^*$.
Following the suggestions of \cite{ZZZW23,LG2025}, we set $2\lambda_1
=\|\bm{\Delta}_{(s)}^\diamondsuit\|_2^2$, where $\|\bm{\Delta}_{(s)}^\diamondsuit\|_2^2$ denotes the $s$th largest value of
$\|\bm\Delta_1^\diamondsuit\|_2^2,\cdots, \|\bm\Delta_J^\diamondsuit\|_2^2$.
Our task here is to select an optimal $s$ ($1\leq s\leq J$) to attain group-wise sparsity. Furthermore, the selection of $s$ is analogous to that of $\lambda$ in $\ell_{1}$ penalties or nonconvex penalties, and we can choose the optimal $s$ by a data driven method later.
\end{remark}

Suppose that $\{\mathcal{S}^k,\mathcal{F}^k,\bm{\mathrm{\beta}}^k\}$ is the solution at the $k$th iteration. Let $\bm{\mathrm{X}}=(\bm {\mathrm{x}}_1,\ldots,\bm {\mathrm{x}}_n)^\mathrm{T}$ and $\bm {\mathrm{Y}}=(\mathrm{y}_1,\ldots,\mathrm{y}_n)^\mathrm{T}$. Recall that $\mathcal{C}= \left\{1,\ldots,J\right\}$ and $\overline{K}(u)=\int_{-\infty}^uK(t)dt$ with $K(\cdot)$ being a kernel function.
We can approximate the results of Proposition \ref{pro.1} iteratively, and summarize it as in Algorithm \ref{alg:1}.

\begin{algorithm}[H]
{\scriptsize
		\caption{Group support detection and root finding algorithm (GSDRFA) via $\ell_{2,0}$-regularized CSQR.}
		\label{alg:1}
		\KwIn{$\bm{\mathrm{X}}$, $\bm{\mathrm{Y}}$, $\{\mathcal{G}_j\}_{j=1}^J$, $s$, $\lambda_1$ and $\mathcal{S}^0$.}

Set $k=0$ and $\mathcal{F}^0= \mathcal{C}\backslash\mathcal{S}^0 $, where $\mathcal{C}= \left\{1,\ldots,J\right\}$.

\While{$k\geq 0$}{
		
		 Obtain $\left(\alpha^k,\bm\beta_{{\mathcal{S}}^k}^k\right)= \mathop {\arg\min }\limits_{\alpha,\bm\beta_{{\mathcal{S}}^k}} \widetilde{\mathcal{L}}_{n,h}\left(\alpha, \bm\beta_{{\mathcal{S}}^k}\right)$.

Set $\bm\beta^k=\left(\left(\bm\beta_{\mathcal{S}^k}^k\right)^{\mathrm{T}},\left(\bm\beta_{\mathcal{F}^k}^k\right)^{\mathrm{T}}\right)^{\mathrm{T}}$ with $\bm\beta_{\mathcal{F}^k}^k=\bm 0$.

\For{$j=1,\ldots,J$,}{
Compute
$\bm{\mathcal{H}}_{\mathcal{G}_j}^k=\frac{1}{n}\sum_{i=1}^{n}K_h\left(-r_{i\mathcal{G}_{j}}^k(\bm\beta_{\mathcal{G}_j}^k)\right)\bm{\mathrm{x}}_{i\mathcal{G}_j}\bm{\mathrm{x}}_{i\mathcal{G}_j}^{\mathrm{T}}$
with $r_{i\mathcal{G}_{j}}^k(\bm\beta_{\mathcal{G}_j}^k)=\mathrm{y}_i-\alpha^k-\bm{\mathrm{x}}_{i}^{\mathrm{T}}\bm{\mathrm{\beta}}^k$.

\If {$j \in \mathcal{F}^k$}{
${\bm d }_{\mathcal{G}_j}^k=-\frac{1}{n}\left(\bm{\mathcal{H}}_{\mathcal{G}_j}^k \right)^{-1}\sum_{i =1}^{n}\left(\overline{K}\left( -r_{i\mathcal{G}_{j}}^k(\bm\beta_{\mathcal{G}_j}^k)/h\right)-\tau\right)\bm{\mathrm{x}}_{i\mathcal{G}_j}$. }

\If {$j \in \mathcal{S}^k$}{
${\bm d }_{\mathcal{G}_j}^k=\bm 0$.
}

Compute $\bm\Delta_j^k=(\bm{\mathcal{H}}_{\mathcal{G}_j}^k)^{1/2} \left(\bm\beta_{\mathcal{G}_j}^k+{\bm d }_{\mathcal{G}_j}^k\right)$.
}
        Update $\mathcal{S}^{k+1}=\{j\in \mathcal{C}:\|\bm\Delta_j^k\|_2^2 \geq \|\bm\Delta_{(s)}^k\|_2^2\}$. Set $\mathcal{F}^{k+1}=\mathcal{C}\backslash\mathcal{S}^{k+1}$.
		
\If {$\mathcal{S}^{k+1}=\mathcal{S}^k$}{
		Stop and denote $\hat{\mathcal{S}}=\mathcal{S}^{k+1}$ and $\hat{\mathcal{F}}=\mathcal{F}^{k+1}$.
	     	}
Set $k\leftarrow k+1$.

		}
	  \KwOut{$\hat{\mathcal{S}}$, $\hat{\mathcal{F}}$, $\hat\alpha$ and $\hat{\bm\beta}=\left((\hat{\bm\beta}_{\hat{\mathcal{S}}})^\mathrm{T},(\hat{\bm\beta}_{\hat{\mathcal{F}}})^\mathrm{T}\right)^\mathrm{T}$, where $\left(\hat\alpha,\hat{\bm\beta}_{\hat{\mathcal{S}}}\right)= \mathop {\arg\min }\limits_{\alpha,\bm\beta_{\hat{\mathcal{S}}}} \widetilde{\mathcal{L}}_{n,h}\left(\alpha, \bm\beta_{\hat{\mathcal{S}}}\right)$ and $\hat{\bm\beta}_{\hat{\mathcal{F}}}=\bm 0$.}
}
\end{algorithm}

\begin{remark}\label{rm1}
In Algorithm \ref{alg:1}, the initial active set $\mathcal{S}^0$ can be obtained via the regularized quantile regression with classical group penalties. This step is readily implementable using the function ``rq.group.pen" with the group adaptive Lasso penalty in the R package ``rqPen" \citep{PW2015}. Moreover, the minimizer of the objective function $\widetilde{\mathcal{L}}_{n,h}\left(\cdot, \cdot\right)$ in Algorithm \ref{alg:1} can be efficiently computed by invoking the function ``conquer" in the R package ``conquer" \citep{HPTZ2023}. Further, Algorithm \ref{alg:1} will be terminated if the iterative active set does not change, i.e., $\mathcal{S}^{k+1}=\mathcal{S}^{k}$.
Algorithm \ref{alg:1} typically terminates within a finite number of iterations in practice.
\end{remark}

It is noteworthy that the outputs of Algorithm \ref{alg:1} depend on the size of active groups $s$, which is typically unknown. To explicitly highlight this dependence on $s$, we rewrite the outputs as $\hat{\mathcal{S}}(s)$, $\hat{\mathcal{F}}(s)$, $\hat\alpha(s)$ and $\hat{\bm\beta}(s)$. Selection of $s$ is somewhat analogous to the choice of $\lambda_1$. However, determining $s$ is no more challenging than determining $\lambda_1$, as $\lambda_1$ is a continuous parameter whereas $s$ is an integer taking values in $\{1,\ldots,J\}$. Furthermore, $s$ has a clear interpretation as the size of active groups, while the interpretation of $\lambda_1$ is less straightforward.
Thus, selection of $s$ plays a crucial role in practice, and a data-driven procedure should be employed to determine it. To trade-off between goodness of fit and model complexity, we suggest to select the optimal $s$ by the following group Bayesian information criterion
(GBIC)
\begin{align}\label{eq8}
    \text{GBIC}(s) &= \log\left( \mathcal{L}_n\left(\hat\alpha(s),\hat{\bm{\mathrm{\beta}}}(s)\right)\right) + \frac{|\hat{\mathcal{S}}(s)|^\diamond \log J \log\log n}{n},
	\end{align}
where $|\hat{\mathcal{S}}(s)|^\diamond = \sum_{j \in \hat{\mathcal{S}}(s)} p_j$, $\mathcal{L}_n(\cdot)$ is given in \eqref{eq2}, $\hat\alpha(s)$ and $\hat{\bm{\mathrm{\beta}}}(s)$ are the final estimators obtained from Algorithm \ref{alg:1} for a given $s$.
Then we can choose the optimal size of active groups by
\begin{align}\label{eq8*}
    \hat{s} = \underset{s \in \{1,
    \ldots, s_{\max}\}}{\mathop{\arg
     \min}} \ \text{GBIC}(s).
\end{align}
where $s_{\max}$ is the upper bound of $s$. We know that some of active groups may be missed when $s_{\max}$ is smaller than $s^*$ but we also suffer from heavy computational burden when $s_{\max}$ is taken too large.
Following \cite{ZZZW23, LG2025}, we take $s_{\max}=\min\{J,\lfloor n/\log p_n\rfloor\}$, where $\lfloor a\rfloor$ is the largest integer no greater than $a$. Numerical evidence in Section 3 shows that the GBIC works well. In other places, the subscript $s$ in $\hat{\mathcal{S}}(s)$, $\hat{\mathcal{F}}(s)$, $\hat\alpha(s)$ and $\hat{\bm\beta}(s)$ is omitted hereafter, and we still use $\hat{\mathcal{S}}$, $\hat{\mathcal{F}}$, $\hat\alpha$ and $\hat{\bm\beta}$ for the simplicity of notation.

\subsection{Theoretical results of Algorithm \ref{alg:1}}
\label{sect:2.2A}

In this subsection, we investigate the statistical properties of Algorithm \ref{alg:1}, including the group support recovery and the consistency of GBIC. We first introduce some regularity conditions.

\begin{enumerate}
    \item[(C1)] For each $1\leq i\leq n$ and $1\leq k\leq p_n$, $|\mathrm{x}_{ik}|$ is uniformly bounded by a positive constant $c$.

    \item[(C2)] The kernel function $K(\cdot)$ is nonnegative and symmetric about zero and satisfies $\int_{-\infty}^{\infty}K(u)\,du=1$ and $\int_{-\infty}^{\infty}u^2K(u)\,du<\infty$, is uniformly bounded, i.e., $\sup_{u\in\mathbb{R}}K(u)\leq\kappa_u$ for some constant $\kappa_u>0$.

    \item[(C3)] The conditional density $f_{\epsilon|\mathrm{x}}(\cdot)$ of $\epsilon$ given $\mathrm{x}$ satisfies $f_{\epsilon|\mathrm{x}}(0)\geq f_l>0$ and $\sup_{t\in\mathbb R}f_{\epsilon|\mathrm{x}}(t)\leq f_u<\infty$ almost surely. It is Lipschitz continuous in a neighborhood of the origin: $|f_{\epsilon|\mathrm{x}}(u)-f_{\epsilon|\mathrm{x}}(v)|\leq l_0|u-v|$ for all $u,v$ sufficiently close to $0$, almost surely.

    \item[(C4)] There exist constants $0<c_-^*(2s_{\max})\leq c_+^*(2s_{\max})<\infty$ such that for any group subset $\mathcal{A}\subseteq\mathcal{C}$ with $|\mathcal{A}|\leq 2s_{\max}$ and any $\bm{u}\neq\bm{0}$, $c_-^*(2s_{\max})\|\bm{u}\|_2^2\leq n^{-1}\|\bm{\mathrm{x}}_{\mathcal{A}}\bm{u}\|_2^2\leq c_+^*(2s_{\max})\|\bm{u}\|_2^2$. The same sparse-eigenvalue bounds hold for the corresponding augmented design including the intercept, that is, for every scalar $a$ and vector $\bm u$, $c_-^*(2s_{\max})(a^2+\|\bm u\|_2^2)\le n^{-1}\sum_{i=1}^n(a+\bm{\mathrm x}_{i\mathcal A}^{\mathrm T}\bm u)^2\le c_+^*(2s_{\max})(a^2+\|\bm u\|_2^2)$.

        \item[(C5)] Let $\omega_s=\max_{\substack{\mathcal A,\mathcal B\subseteq\mathcal C:\,|\mathcal A|,|\mathcal B|\le s,\ \mathcal A\cap\mathcal B=\varnothing}}\ \max_{\bm u\ne\bm0}
    \frac{\left\|n^{-1}\sum_{i=1}^n f_{\epsilon|\bm{\mathrm{x}}_i}(0)\,\bm{\mathrm{x}}_{i\mathcal A}\bm{\mathrm{x}}_{i\mathcal B}^{\top}\bm u\right\|_2}{f_l\|\bm u\|_2}$. Assume that $\omega_{2s_{\max}}\le\kappa$ for a sufficiently small constant $\kappa>0$.

\item[(C6)] (i) The minimum group signal $\vartheta_{\min}=\min\limits_{j\in\mathcal{S}^*}\|\bm{\beta}_{\mathcal{G}_j}^*\|_2$ satisfies
$\vartheta_{\min}^2\gg s_{\max}p_{\max}\{\frac{\log p_n}{n f_l^2}+h^4\}$, where $p_{\max}=\max_{1\leq j\leq J}p_j$. (ii) Assume that $|\mathcal S^0|\le s_{\max}$ and that there exists a sufficiently small fixed constant $\rho_G>0$, independent of $n$, such that
$\mathbb P\!\left\{\left\|\bm\beta^*_{\mathcal G_{\mathcal S^*\setminus\mathcal S^0}}\right\|_2>\rho_G\right\}\to0$, where $\mathcal S^0$ is the initial active-group set supplied to Algorithm~\ref{alg:1}.

\item[(C7)] The bandwidth $h$ satisfies $h\to0$, $nh\to\infty$, and $s_{\max}^{2}p_{\max}\log p_n/(nh)\to0$.

\item[(C8)] $s_{\max}p_{\max}\log p_n/n\to0$, $p_{\max}\log J\log\log n/(n\vartheta_{\min}^2)\to0$, $\log p_n/\{\log J\log\log n\}\to0$, and $nh^4/\{\log J\log\log n\}\to0$.

\end{enumerate}

Condition (C1) is standard in high-dimensional statistics \citep{BC2011,TWZ2022}. Conditions (C2)--(C3) ensure convex smoothing, a bounded local curvature, and a second-order smoothing bias. Together with (C4) and (C7), they yield uniform positive definiteness and concentration of the smoothed Hessian. Condition (C4) is the group sparse Riesz condition (GSRC), which controls the design eigenvalues on sparse group subspaces and their augmentation by the unpenalized intercept. Condition (C5) is a standard group incoherence (sparse-orthogonality) condition expressed in the Hessian metric. Once group sparsity is controlled, distinct groups, and the intercept, are only weakly cross-correlated, which is far weaker than requiring pairwise orthogonality, and it guarantees geometric contraction of the group-level support update. Condition (C6)(i) requires the weakest active-group signal to
dominate the stochastic estimation error and smoothing bias.
Condition (C6)(ii) controls the size of the initial active set
and the total signal omitted by that set, supporting the local
convergence analysis. It allows both omitted active groups and
included inactive groups, provided that the omitted signal is
sufficiently small with high probability. Since $\rho_G$ is fixed, the omitted signal need not vanish, and no consistent initial
parameter estimator is required. Condition (C7) gives the bandwidth rate needed for uniform Hessian concentration, whereas (C8) contains only the additional growth, smoothing-bias, and signal-to-penalty requirements needed for GBIC consistency.

\begin{theorem}\label{thm:1}
Let $\hat{\mathcal S}(s)$ be the output of Algorithm~\ref{alg:1} with the input $s\ge s^*$.
Under Conditions~(C1)--(C7), we have
\[\mathbb P\{\hat{\mathcal S}(s)\supseteq\mathcal S^*\}\ge1-\delta_n,\]
where
$\delta_n\le C s_{\max}p_{\max}J^{-c_0s_{\max}}+C p_n^{-c_1}+\mathbb P(\mathcal E_{G,0}^c)$
for constants $c_0,c_1,C>0$ and $\mathcal E_{G,0}=\left\{|\mathcal S^0|\le s_{\max},\
\|\bm\beta^*_{\mathcal G_{\mathcal S^*\setminus\mathcal S^0}}\|_2\le \rho_G\right\}$ with $\rho_G$ given in Condition (C6).
%In particular, Condition~(C6) implies $\delta_n\to0$. If $s=s^*$, then $\mathbb P\{\hat{\mathcal S}(s)=\mathcal S^*\}\ge1-\delta_n$.
\end{theorem}

\begin{theorem}\label{thm:2}
Let $\hat{\mathcal S}=\hat{\mathcal S}(\hat s)$ be the output of Algorithm~\ref{alg:1} with the input $s= \hat s$, where $\hat s$ is determined by \eqref{eq8*}. Under Conditions~(C1)--(C8), we have
\[\mathbb P(\hat s=s^*,\ \hat{\mathcal S}=\mathcal S^*)\ge1-\delta_{G,n},\]
where $\delta_{G,n}\le  Cp_n^{-c_1}+\mathbb P(\mathcal E_{G,0}^c)
+C\exp\{-cn\vartheta_{\min}^2/(s_{\max}p_{\max})\}
+C\exp\{-c\log J\log\log n\}+C s_{\max}p_{\max}J^{-c_0s_{\max}}$
for constants $c_0,c_1,c,C>0$, with $\mathcal E_{G,0}$ defined in Theorem~\ref{thm:1}.
%Consequently, $\delta_{G,n}\to0$ under Conditions~(C1)--(C8).
\end{theorem}

\begin{remark}\label{rm1A}
Under Conditions~(C1)--(C8), it is easy to check $\delta_n\to0$ and $\delta_{G,n}\to0$. Consequently, Theorem~\ref{thm:1} implies that $\hat{\mathcal S}(s)$ can recover the true subset $S^*$ with high probability
only if the given size $s$ is greater than the true active group size $s^*$. Theorem~\ref{thm:2} further shows that GBIC consistently selects $s^*$ and therefore yields exact group recovery.
\end{remark}

\subsection{$\ell_{0}$-regularized CSQR for Within-group Sparsity}

When a group is selected as the active group, all variables in the group are estimated as non-zero, which is insufficient for some applications like gene pathway analysis. For instance, in the gene-set selection problem, a biological pathway may be associated with a specific biological process, but this does not necessarily imply that all genes within the pathway are related to the process. Biomedical workers therefore seek to not only effectively eliminate irrelevant pathways but also identify key genes within the relevant ones. Consequently, we can further eliminate the irrelevant variables from the active groups selected by Algorithm \ref{alg:1} to achieve within-group sparsity.

Let's recall some notations.
Recall that $\mathcal{G}_{\hat{\mathcal{S}}}=\bigcup_{j\in \hat{\mathcal{S}}}\mathcal{G}_j$, $|\hat{\mathcal{S}}|^\diamond=\sum_{j \in\hat{\mathcal{S}}}p_j$, $\bm\beta_{\hat{\mathcal{S}}}=\big(\bm\beta_{\mathcal{G}_j}^{\mathrm{T}},j \in \hat{\mathcal{S}}\big)^{\mathrm{T}} \in \mathbb{R}^{|\hat{\mathcal{S}}|^\diamond}$ with $\bm\beta_{\mathcal{G}_j}=\big(\beta_{j'},j' \in \mathcal{G}_j\big)^{\mathrm{T}} \in \mathbb{R}^{p_j}$ and $\bm {\mathrm{x}}_{i\hat{\mathcal{S}}}=\big(\bm {\mathrm{x}}_{i\mathcal{G}_j}^{\mathrm{T}}, j \in \hat{\mathcal{S}} \big)^{\mathrm{T}}\in \mathbb{R}^{|\hat{\mathcal{S}}|^\diamond}$ with $\bm {\mathrm{x}}_{i\mathcal{G}_j}=\big({\mathrm{x}}_{ik},k\in \mathcal{G}_j \big)^{\mathrm{T}} \in \mathbb{R}^{p_j}$, where ${\mathrm{x}}_{ik}$ is the $k$th component of $\bm {\mathrm{x}}_i$. By the definition of $\mathcal{G}_{\hat{\mathcal{S}}}$, it is easy to verify $\bm\beta_{\hat{\mathcal{S}}}=\big(\bm\beta_{\mathcal{G}_j}^{\mathrm{T}},j \in \hat{\mathcal{S}}\big)^{\mathrm{T}}=\big(\beta_{j'},j'\in\mathcal{G}_{\hat{\mathcal{S}}}\big)^{\mathrm{T}}
\triangleq \bm\beta_{\mathcal{G}_{\hat{\mathcal{S}}}}$. Without loss of generality, we assume $\mathcal{G}_{\hat{\mathcal{S}}}\triangleq\{s_1,\cdots,s_{|\hat{\mathcal{S}}|^\diamond}\}\subset \left\{1,\cdots,p_n\right\}$ for the convenience of presentation, implying $\bm\beta_{\mathcal{G}_{\hat{\mathcal{S}}}}=\big(\beta_{s_1},\cdots,\beta_{s_{|\hat{\mathcal{S}}|^\diamond}}\big)^{\mathrm{T}}$.
Since we do not use a group structure in the subsequent analysis, we hereafter use the notation $\bm\beta_{\mathcal{G}_{\hat{\mathcal{S}}}}$ (in lieu of $\bm\beta_{\hat{\mathcal{S}}}$) for notational clarity. It should be emphasized, however, that $\bm{\beta}_{\mathcal{G}_{\hat{\mathcal{S}}}}$ is exactly equivalent to $\bm{\beta}_{\hat{\mathcal{S}}}$ ($\bm{\beta}_{\mathcal{G}_{\hat{\mathcal{S}}}}=\bm{\beta}_{\hat{\mathcal{S}}}$).

Note that $\bm\beta_{\mathcal{G}_{\hat{\mathcal{S}}}}$ might be sparse, implying that some components in $\bm\beta_{\mathcal{G}_{\hat{\mathcal{S}}}}$ are zeros. Although Algorithm \ref{alg:1} can eliminate irrelevant groups (achieve the goal of group-wise sparsity), it results in all components of $\bm\beta_{\mathcal{G}_{\hat{\mathcal{S}}}}$ being estimated as non-zeros, and hence can not achieve within-group sparsity.
Therefore, for a given $\hat{\mathcal{S}}$, we suggest to re-estimate $\bm\beta_{\mathcal{G}_{\hat{\mathcal{S}}}}$ by minimizing the following $\ell_0$-regularized CSQR
\begin{align}\label{eq:10}
\min_{\alpha,\bm\beta_{\mathcal{G}_{\hat{\mathcal{S}}}}}
\widetilde{\mathcal{L}}_{n,h,\lambda_2}(\alpha,\bm\beta_{\mathcal{G}_{\hat{\mathcal{S}}}}),
\end{align}
where $\widetilde{\mathcal{L}}_{n,h,\lambda_2}(\alpha,\bm\beta_{\mathcal{G}_{\hat{\mathcal{S}}}})
=\widetilde{\mathcal{L}}_{n,h}(\alpha,\bm\beta_{\mathcal{G}_{\hat{\mathcal{S}}}})+ \lambda_2 \|\bm\beta_{\mathcal{G}_{\hat{\mathcal{S}}}}\|_{0}$, $\|\bm\beta_{\mathcal{G}_{\hat{\mathcal{S}}}}\|_{0}
=\sum_{j =1}^{{|\hat{\mathcal{S}}|^\diamond}}  \mathbb{I} \{|\beta_{s_j}| \neq 0\}$ and $\lambda_2>0$ controls the within-group sparsity level.

Similar to \eqref{eq6}, we say that
$(\alpha^\natural,
\bm\beta_{\mathcal{G}_{\hat{\mathcal{S}}}}^\natural)$
is a coordinate-wise minimizer of \eqref{eq:10} if
\[
\alpha^\natural \in
\arg\min_{\alpha}
\widetilde{\mathcal L}_{n,h}
\left(\alpha,
\bm\beta_{\mathcal{G}_{\hat{\mathcal{S}}}}^\natural\right),
\]
and, for every $j=1,\ldots,|\hat{\mathcal S}|^\diamond$,
\[
\begin{aligned}
\beta_{s_j}^\natural
\in\arg\min_{\beta_{s_j}}
\widetilde{\mathcal L}_{n,h,\lambda_2}
\big(&\alpha^\natural,
\beta_{s_1}^\natural,\ldots,
\beta_{s_{j-1}}^\natural,\beta_{s_j},
\beta_{s_{j+1}}^\natural,\ldots,
\beta_{s_{|\hat{\mathcal S}|^\diamond}}^\natural
\big).
\end{aligned}
\]
Furthermore, we need to define $\Delta^{\natural}_{s_j}=\sqrt{\varsigma_{s_j}^\natural} \left(\beta_{s_j}^\natural+ \xi_{s_j}^\natural\right)$, $\varsigma_{s_j}^\natural = \frac{{\partial ^2 {{\widetilde{\cal L}}_{n,h}}\left(\alpha, \bm\beta_{\mathcal{G}_{\hat{\mathcal{S}}}} \right)}}{ {\partial \beta_{s_j}^2 }}\left| {_{\alpha=\alpha^\natural,\bm\beta_{\mathcal{G}_{\hat{\mathcal{S}}}}  = \bm\beta_{\mathcal{G}_{\hat{\mathcal{S}}}}^\natural}} \right.= \frac{1}{n}\sum_{i=1}^n \mathrm{x}_{is_j}^2K_h\left(-r_{is_j}^\natural(\beta_{s_j}^\natural)\right)$,
$r_{is_j}^\natural(\beta_{s_j})=\mathrm{y}_i-\mathrm{x}_{is_j}\beta_{s_j}-\alpha^\natural-\sum_{j'=1,j'\neq j}^{|\hat{\mathcal{S}}|^\diamond}\mathrm{x}_{is_{j'}}\beta_{s_{j'}}^\natural$ and $\xi_{s_j}^\natural=-(\varsigma_{s_j}^\natural)^{-1}\frac{\partial {{\widetilde{\cal L}}_{n,h}}\left(\alpha, \bm\beta_{\mathcal{G}_{\hat{\mathcal{S}}}} \right)}{\partial \beta_{s_j}}\left| {_{\alpha=\alpha^\natural,\bm\beta_{\mathcal{G}_{\hat{\mathcal{S}}}}  = \bm\beta_{\mathcal{G}_{\hat{\mathcal{S}}}}^\natural}} \right. = -\frac{1}{n}(\varsigma_{s_j}^\natural)^{-1}\sum_{i=1}^{n}\mathrm{x}_{is_j}\left(\overline{K}(-r_{is_j}^\natural(\beta_{s_j}^\natural)/h)-\tau\right)$.
For the sake of illustrating the sparse solution corresponding to \eqref{eq:10}, we establish a practical necessary and sufficient condition in Proposition \ref{pro.2}.

\begin{proposition}
\label{pro.2}
{Assume that, for each $s_j\in\mathcal G_{\hat{\mathcal S}}$,}
$\varsigma_{s_j}^{\natural}>0$ and
$0\leq
\frac{\partial^2\widetilde{\mathcal L}_{n,h}
\left(\alpha^\natural,\bm\beta_{\mathcal G_{\hat{\mathcal S}}}\right)}
{\partial\beta_{s_j}^2}
\leq\varsigma_{s_j}^{\natural}$
for
$\beta_{s_j}\in[0,\beta_{s_j}^{\natural}]
\cup[0,\xi_{s_j}^{\natural}]$,
where $\beta_k=\beta_k^\natural$ for all
$k\in\mathcal G_{\hat{\mathcal S}}$ with $k\neq s_j$.
{If $(\alpha^\natural,\bm\beta_{\mathcal{G}_{\hat{\mathcal{S}}}}^{\natural})$ is a coordinate-wise minimizer of \eqref{eq:10} with respect to the intercept and each coordinate, then it satisfies}
\begin{align}\label{eq61}
\left\{ \begin{array}{l}
{\mathcal A^{\natural}=\{s_j\in\mathcal G_{\hat{\mathcal S}}:|\Delta_{s_j}^{\natural}|\ge\sqrt{2\lambda_2}\}}\\
{\mathcal I^{\natural}=\{s_j\in\mathcal G_{\hat{\mathcal S}}:|\Delta_{s_j}^{\natural}|<\sqrt{2\lambda_2}\}}\\
\left(\alpha^\natural,\bm\beta_{{\mathcal{A}}^\natural}^\natural\right) =\mathop {\arg\min }\limits_{\alpha,\bm\beta_{{\mathcal{A}}^\natural}} \widetilde{\mathcal{L}}_{n,h}\left(\alpha, \bm\beta_{{\mathcal{A}}^\natural}\right)\\
\bm\beta_{\mathcal{I}^{\natural}}^{\natural} =\bm 0
\end{array} \right.
\end{align}
where $\widetilde{\mathcal{L}}_{n,h}\left(\alpha,\bm\beta_{{\mathcal{A}}^\natural} \right)=\frac{1}{n}\sum_{i =1}^{n}\ell_h\left(\mathrm{y}_i-\alpha-\bm {\mathrm{x}}_{i{{\mathcal{A}}^\natural}}^{\mathrm{T}}\bm\beta_{{\mathcal{A}}^\natural}\right)$, $\bm{\mathrm{\beta}}_{{\mathcal{A}}^\natural}^\natural=\left(\mathrm{\beta}_{s_j}^\natural, s_j \in{\mathcal{A}}^\natural\right)^{\mathrm{T}}$ and $\bm{\mathrm{\beta}}_{{\mathcal{I}}^\natural}^\natural=\left(\mathrm{\beta}_{s_j}^\natural,s_j \in{\mathcal{I}}^\natural\right)^{\mathrm{T}}$.
{Conversely, if $(\alpha^\natural,\bm\beta_{\mathcal G_{\hat{\mathcal S}}}^{\natural})$ satisfies \eqref{eq61}, then $(\alpha^\natural,\bm\beta_{\mathcal G_{\hat{\mathcal S}}}^{\natural})$ is a local minimizer of \eqref{eq:10}.}
\end{proposition}

It is important to distinguish the notation in the two propositions. In Proposition~\ref{pro.1}, $\mathcal S^\diamondsuit\subseteq\mathcal C$ is a set of \emph{group indices}; in Proposition~\ref{pro.2}, $\mathcal A^\natural\subseteq\mathcal G_{\hat{\mathcal S}}$ is a set of \emph{individual variable indices} within the groups retained by the first stage. Hence $\mathcal S^\diamondsuit$ and $\mathcal A^\natural$ describe sparsity at different levels and should not be interpreted as the same type of support. Neither should be confused with the population supports $\mathcal S^*$ and $\mathcal A^*$ used in the asymptotic theory.

Proposition \ref{pro.2} provides a principled thresholding rule to prune irrelevant variables within the active group $\mathcal{G}_{\hat{\mathcal{S}}}$, achieving within-group sparsity. Its proof is given in the Supplementary Material.
Regarding the selection of $\lambda_2$, we adopt a similar strategy as used for $\lambda_1$. We set $\sqrt{2\lambda_2}=|\Delta_{(r)}^{\natural}|$, where $|\Delta_{(r)}^{\natural}|$ denotes the $r$th largest value of $|\Delta_{s_1}^{\natural}|,\cdots,|\Delta_{s_{|\hat{\mathcal{S}}|^\diamond}}^{\natural}|$.
%When the $r$th and $(r+1)$st scores are distinct, Proposition \ref{pro.2} selects exactly the top $r$ variables; under a boundary tie, Algorithm \ref{alg:2} follows its stated $\geq$ updating rule.
The parameter $r$ determines the level of within-group sparsity, and we discuss how to determine it later. Suppose that $\{\mathcal{A}^k,\mathcal{I}^k,\bm{\beta}_{\mathcal{G}_{\hat{\mathcal{S}}}}^k\}$ is the current solution at the $k$th iteration, we can approximate the results of Proposition \ref{pro.2} iteratively, and summarize it as in Algorithm \ref{alg:2}.

\begin{algorithm}[H]
{\scriptsize
	\caption{Sparse support detection and root finding algorithm (SSDRFA) via $\ell_{0}$-regularized CSQR.}
	\label{alg:2}
	\KwIn{$\bm{\mathrm{X}}$, $\bm{\mathrm{Y}}$, $\hat{\mathcal{S}}$, $r$ and $\mathcal{A}^{0}$.}

Set $k=0$, $\mathcal{G}_{\hat{\mathcal{S}}}=\bigcup_{j\in \hat{\mathcal{S}}}\mathcal{G}_j=\{s_1,\cdots,s_{|\hat{\mathcal{S}}|^\diamond}\}$ and $\mathcal{I}^{0}= \mathcal{G}_{\hat{\mathcal{S}}}\backslash \mathcal{A}^{0}$ .

\While{$k\geq 0$}{

Obtain $\left(\alpha^{k},\bm{\mathrm{\beta}}_{\mathcal{A}^k}^{k}\right)=\mathop {\arg\min}\limits_{\alpha,\bm\beta_{{\mathcal{A}}^k}} \widetilde{\mathcal{L}}_{n,h}\left(\alpha, \bm\beta_{{\mathcal{A}}^k}\right)$.

Set $\bm{\mathrm{\beta}}_{\mathcal{G}_{\hat{\mathcal{S}}}}^{k}=\left(\left(\bm{\mathrm{\beta}}_{\mathcal{A}^k}^{k}\right)^{\mathrm{T}},\left(\bm{\mathrm{\beta}}_{\mathcal{I}^k}^{k}\right)^{\mathrm{T}}\right)^{\mathrm{T}}$ with $\bm{\mathrm{\beta}}_{\mathcal{I}^k}^{k}=\bm 0$.

	 \For{$j=1,\cdots,|\hat{\mathcal{S}}|^\diamond$}{
		Compute $\varsigma_{s_j}^k = \frac{1}{n}\sum\limits_{i=1}^n \mathrm{x}_{is_j}^2K_h\left(-r_{is_j}^k(\beta_{s_j}^k)\right)$ with $r_{is_j}^k(\beta_{s_j}^k)=\mathrm{y}_i-\alpha^k-
\bm{\mathrm{x}}_{i\mathcal{G}_{\hat{\mathcal{S}}}}^{\mathrm{T}}\bm\beta_{\mathcal{G}_{\hat{\mathcal{S}}}}^k$.

\If {$s_j \in \mathcal{I}^k$}{
$\xi_{s_j}^k= -\frac{1}{n}(\varsigma_{s_j}^k)^{-1}\sum_{i=1}^{n}\mathrm{x}_{is_j}\left(\overline{K}(-r_{is_j}^k(\beta_{s_j}^k)/h)-\tau\right)$.
}

\If {$s_j \in \mathcal{A}^k$}{
$\xi_{s_j}^k =0$.
}

Compute $\Delta^k_{s_j}=\sqrt{\varsigma_{s_j}^k} \left(\beta_{s_j}^k+ \xi_{s_j}^k\right)$.
	}

Update $\mathcal{A}^{k+1}=\{s_j\in \mathcal{G}_{\hat{\mathcal{S}}}:|\Delta^k_{s_j}| \geq |\Delta^k_{(r)}|, j=1,\cdots,|\hat{\mathcal{S}}|^\diamond \}$. Set $\mathcal{I}^{k+1}=\mathcal{G}_{\hat{\mathcal{S}}}\backslash{\mathcal{A}^{k+1}}$.

\If {$\mathcal{A}^{k+1}=\mathcal{A}^k$}{
			Stop and denote $\bar{\mathcal{A}}=\mathcal{A}^{k+1}$ and $\bar{\mathcal{I}}=\mathcal{I}^{k+1}$.
		}
Set $k\leftarrow k+1$.
}
	\KwOut{$\bar{\mathcal{A}}$, $\bar{\mathcal{I}}$, $\bar\alpha$ and $\bar{\bm\beta}_{\mathcal{G}_{\hat{\mathcal{S}}}}=\left((\bar{\bm\beta}_{\bar{\mathcal{A}}})^\mathrm{T},(\bar{\bm\beta}_{\bar{\mathcal{I}}})^\mathrm{T}\right)^\mathrm{T}$, where $\left(\bar\alpha,\bar{\bm\beta}_{\bar{\mathcal{A}}}\right)= \mathop {\arg\min }\limits_{\alpha,\bm\beta_{\bar{\mathcal{A}}}} \widetilde{\mathcal{L}}_{n,h}\left(\alpha, \bm\beta_{\bar{\mathcal{A}}}\right)$ and $\bar{\bm\beta}_{\bar{\mathcal{I}}}=\bm 0$.}
}
\end{algorithm}

\begin{remark}\label{rm2}
Once the active group set $\hat{\mathcal{S}}$ is obtained by implementing Algorithm \ref{alg:1}, we use the function ``rq.pen" in the R package ``rqPen" with the adaptive Lasso penalty to obtain an initial active set $\mathcal{A}^0$ for Algorithm \ref{alg:2}.
Moreover, the dimension of variables involved in Algorithm \ref{alg:2} is $|\hat{\mathcal{S}}|^\diamond$ (far smaller than the original dimension $p_n$), as Algorithm \ref{alg:1} achieves effective dimension reduction by filtering out irrelevant groups. As a result, Algorithm \ref{alg:2} demonstrates exceptional computational efficiency.
\end{remark}

Following the aforementioned selection criterion for the group-wise sparsity parameter $s$, we use an analogous criterion to select the within-group sparsity parameter $r$. Specifically, we re-write the outputs of Algorithm \ref{alg:2} as $\bar{\mathcal{A}}(r)$, $\bar{\mathcal{I}}(r)$, $\bar\alpha(r)$ and $\bar{\bm\beta}_{\mathcal{G}_{\hat{\mathcal{S}}}}(r)$ to explicitly indicate that these outputs rely on $r$.
Then, we choose the optimal $r$ via the following Bayesian Information Criterion (BIC)
    \begin{align}\label{eq15}
    \text{BIC}(r) &= \log\left( \mathcal{L}_n\left(\bar\alpha(r),\bar{\bm{\mathrm{\beta}}}(r)\right)\right) + \frac{|\bar{\mathcal{A}}(r)| \log (|\hat{\mathcal{S}}|^\diamond) \log\log n}{n},
	\end{align}
where $|\bar{\mathcal{A}}(r)|$ is the number
of elements in $\bar{\mathcal{A}}(r)$, $|\hat{\mathcal{S}}|^\diamond=\sum_{j\in\hat{\mathcal{S}}}p_j$, $\bar\alpha(r)$ and $\bar{\bm{\mathrm{\beta}}}(r)$ are the final estimators obtained from Algorithm \ref{alg:2} for a given $r$.
Then, the optimal within-group sparsity level $\bar{r}$ is determined by
   \begin{align}\label{eq15*}
	\bar{r} = \underset{r \in \{1, \ldots, r_{\max}\}}{\mathop{\arg \min}} \ \text{BIC}(r).
    \end{align}	
where $r_{\max}=\min\{|\hat{\mathcal{S}}|^\diamond,\lfloor n/\log(|\hat{\mathcal{S}}|^\diamond)\rfloor\}$ is the upper bound for $r$ and $\lfloor a\rfloor$ is the largest integer no greater than $a$.

\subsection{Theoretical results of Algorithms \ref{alg:2} and \ref{alg:3}}
\label{sect:2.2AAA}

The candidate group set $\hat{\mathcal S}$ produced by
Algorithm~\ref{alg:1} is random. The first-stage theory guarantees
$\mathcal S^*\subseteq\hat{\mathcal S}$ with high probability, with
equality when the true group model size is used. We therefore account
for first-stage selection uncertainty when establishing the
second-stage recovery and model-selection results.

\begin{enumerate}
\item[(C9)]
(i) The minimum active-variable signal satisfies
$
(\beta_{\min}^*)^2
\gg
r_{\max}
\left\{
\frac{\log p_n}{nf_l^2}+h^4
\right\}
+
\frac{\log(s_{\max}p_{\max})\log\log n}{n},
$
and
$
\frac{\log p_n}
{\log(|\mathcal S^*|^\diamond)\log\log n}
\to0,~
\frac{nh^4}
{\log(|\mathcal S^*|^\diamond)\log\log n}
\to0,
$
where
$\beta_{\min}^*=\min_{k\in\mathcal A^*}|\beta_k^*|$.

(ii) The initial active-variable set $\mathcal A^0$ supplied to
Algorithm~\ref{alg:2} satisfies $|\mathcal A^0|\le r_{\max}$.
Moreover, there exists a sufficiently small fixed constant
$\rho_V>0$, independent of $n$, such that
$
\mathbb P\!\left\{
\left\|
\bm\beta^*_{\mathcal A^*\setminus\mathcal A^0}
\right\|_2>\rho_V
\,\middle|\,
\mathcal S^*\subseteq\hat{\mathcal S}
\right\}
\to0.
$

\item[(C10)]
Assume that $\omega^V_{2r_{\max}}\le\kappa$ for a sufficiently
small constant $\kappa>0$, where
$
\omega_r^V
=
\max_{\substack{
\mathcal A,\mathcal B\subseteq\mathcal G_{\hat{\mathcal S}}\\
|\mathcal A|,|\mathcal B|\le r,\;
\mathcal A\cap\mathcal B=\varnothing
}}
\ \max_{\bm u\ne\bm0}
\frac{
\left\|
n^{-1}\sum_{i=1}^n
f_{\epsilon|\mathrm{x}_i}(0)
\,\mathrm{x}_{i\mathcal A}
\mathrm{x}_{i\mathcal B}^{\top}\bm u
\right\|_2
}{
f_l\|\bm u\|_2
}.
$
\end{enumerate}

Condition~(C9)(i) requires the minimum active-variable signal to
dominate the second-stage stochastic error, smoothing bias, and
BIC penalty, together with the growth restrictions needed for
model-selection consistency.
Condition~(C9)(ii) controls the size of the initial active-variable
set and the total signal omitted by that set, conditional on
first-stage sure screening.
Since $\rho_V$ is fixed, the omitted signal need not be $o(1)$,
and no consistent initial parameter estimator is required.
Condition~(C10) is the variable-level counterpart of
Condition~(C5). It controls the density-weighted dependence between
disjoint variable subsets within the selected candidate set and,
together with the other conditions, supports the geometric
contraction analysis of Algorithm~\ref{alg:2}.

The following results establish second-stage support recovery and
BIC consistency while explicitly accounting for the probability
of first-stage screening failure.

\begin{theorem}\label{thm:3}
Let $\bar{\mathcal A}(r)$ be the output of
Algorithm~\ref{alg:2} with input $r\ge r^*$.
Under Conditions~(C1)--(C10),
\[
\mathbb P\{\bar{\mathcal A}(r)\supseteq\mathcal A^*\}
\ge
1-\delta_{G,n}-\delta_{V,n},
\]
where $\delta_{G,n}$ is defined in
Theorem~\ref{thm:2}, and
$
\delta_{V,n}
\le
\frac{
C s_{\max}p_{\max}J^{-c_0s_{\max}}
+
Cp_n^{-c_1}
}{
1-\delta_{G,n}
}
+
\mathbb P\!\left(
\mathcal E_{V,0}^c
\,\middle|\,
\mathcal S^*\subseteq\hat{\mathcal S}
\right)
$
for constants $c_0,c_1,C>0$. Here,
$
\mathcal E_{V,0}
=
\left\{
|\mathcal A^0|\le r_{\max},\;
\left\|
\bm\beta^*_{\mathcal A^*\setminus\mathcal A^0}
\right\|_2\le\rho_V
\right\},
$
with $\rho_V$ specified in Condition~(C9).
\end{theorem}

\begin{theorem}\label{thm:4}
Let $\bar r$ be selected by \eqref{eq15*}, and let
$\bar{\mathcal A}=\bar{\mathcal A}(\bar r)$ be the output
of Algorithm~\ref{alg:2} with input $r=\bar r$.
Under the conditions of Theorem~\ref{thm:3},
\[
\mathbb P\{
\bar r=r^*,\;
\bar{\mathcal A}=\mathcal A^*
\}
\ge
1-\delta_{G,n}-\delta_{V,n}-\eta_{V,n},
\]
where
$
\eta_{V,n}
\le
\frac{
C\exp\!\left\{
-c n(\beta_{\min}^*)^2/r_{\max}
\right\}
+
C\exp\!\left\{
-c\log(|\mathcal S^*|^\diamond)\log\log n
\right\}
}{
1-\delta_{G,n}
}
$
for constants $c,C>0$, and $\delta_{G,n}$ and
$\delta_{V,n}$ are defined in Theorems~\ref{thm:2}
and~\ref{thm:3}, respectively.
\end{theorem}

\begin{remark}
Theorems~\ref{thm:3} and~\ref{thm:4} account explicitly
for first-stage selection uncertainty.
The second stage requires only
$\mathcal S^*\subseteq\hat{\mathcal S}$, allowing the
candidate set to contain inactive groups.
As the failure bounds vanish, Algorithm~\ref{alg:2}
retains all active variables with probability tending to one,
and the BIC criterion consistently selects the true
within-group model size, yielding exact recovery of
$\mathcal A^*$.
\end{remark}

In order to present the whole sparse-group selection process,
we integrate Algorithms \ref{alg:1}--\ref{alg:2} and two information criteria (GBIC and BIC) to form the
following final process.

\begin{algorithm}[H]
{\scriptsize
	\caption{Sparse-group selection algorithm for high dimensional QR (SGSA-HDQR).}
	\label{alg:3}
	\KwIn{$\bm{\mathrm{X}}$, $\bm{\mathrm{Y}}$ and $\{\mathcal{G}_j\}_{j=1}^J$.}

Obtain an initial set of active groups $\mathcal{S}^0$ from $\mathcal{C}$ by using the function ``rq.group.pen" in the R package ``rqPen".

\For{$s=1,\cdots,s_{\max}$}{
$\left(\hat{\mathcal{S}}(s), \hat{\mathcal{F}}(s), \hat\alpha(s),\hat{\bm\beta}(s)\right)={\rm Algorithm}~ \ref{alg:1}\big(\bm{\mathrm{X}}, \bm {\mathrm{Y}}, \left\{\mathcal{G}_j\right\}_{j=1}^J, s, \mathcal{S}^0\big)$

Computing ${\rm GBIC}\left(s\right)$ given in \eqref{eq8}.
}

Obtain the optimal $\hat s$ by \eqref{eq8*}, and set $\hat{\mathcal{S}}\triangleq \hat{\mathcal{S}}(\hat s)$, $\hat{\mathcal{F}}=\mathcal{C}\setminus\hat{\mathcal{S}}$, $\mathcal{G}_{\hat{\mathcal{S}}}=\bigcup_{j\in \hat{\mathcal{S}}}\mathcal{G}_j$ and $\hat{\bm\beta}=\left((\hat{\bm\beta}_{\hat{\mathcal{S}}})^\mathrm{T},(\hat{\bm\beta}_{\hat{\mathcal{F}}})^\mathrm{T}\right)^\mathrm{T}$ with $\hat{\bm\beta}_{\hat{\mathcal{F}}}=\bm 0$ .

Obtain an initial active set $\mathcal{A}^0$ from $\mathcal{G}_{\hat{\mathcal{S}}}$ by using the function ``rq.pen" in the R package ``rqPen".

\For{$r=1,\cdots,r_{\max}$}{
$\left(\bar{\mathcal{A}}(r), \bar{\mathcal{I}}(r), \bar\alpha(r),\bar{\bm\beta}_{\mathcal{G}_{\hat{\mathcal{S}}}}(r)\right)={\rm Algorithm}~ \ref{alg:2}\big(\bm{\mathrm{X}}, \bm {\mathrm{Y}}, \hat{\mathcal{S}},r, \mathcal{A}^0\big)$

Computing ${\rm BIC}\left(r\right)$ given in \eqref{eq15}.
}

Obtain the optimal $\bar r$ by \eqref{eq15*}, and set $\bar{\mathcal{A}}\triangleq\bar{\mathcal{A}}(\bar r)$, $\bar{\mathcal{I}}=\mathcal{G}_{\hat{\mathcal{S}}}\backslash \bar{\mathcal{A}}$,
$\bar\alpha=\bar\alpha(\bar r)$ and $\bar{\bm\beta}_{\mathcal{G}_{\hat{\mathcal{S}}} }\triangleq\bar{\bm\beta}_{\mathcal{G}_{\hat{\mathcal{S}}}}(\bar r)=\left((\bar{\bm\beta}_{\bar{\mathcal{A}}})^\mathrm{T},
(\bar{\bm\beta}_{\bar{\mathcal{I}}})^\mathrm{T}\right)^\mathrm{T}$ with $\bar{\bm\beta}_{\bar{\mathcal{I}}}=\bm 0$.

	\KwOut{$\hat{\mathcal{S}}$, $\hat{\mathcal{F}}$, $\bar{\mathcal{A}}$, $\bar{\mathcal{I}}$, $\bar\alpha$ and $\hat{\bm\beta}=\left((\bar{\bm\beta}_{\mathcal{G}_{\hat{\mathcal{S}}} })^\mathrm{T},(\hat{\bm\beta}_{\hat{\mathcal{F}}})^\mathrm{T}\right)^\mathrm{T}$.}
}
\end{algorithm}

\begin{remark}\label{rm3}
Algorithm \ref{alg:3} is composed of two successive stages. In the first stage, we run Algorithm \ref{alg:1} with the size of active groups determined by GBIC, which allows us to rapidly eliminate irrelevant groups. In the second stage,
we use BIC to further choose the optimal size of active variables in $\mathcal{G}_{\hat{\mathcal{S}}}$, resulting in within-group sparsity. Thus, Algorithm \ref{alg:3} rapidly achieves the goal of sparse-group
selection.
In contrast to traditional single-variable selection methods that identify significant variables directly from the original $p_n$ predictors (e.g., \cite{WWL2012,PW2015,CLMY20,TWZ2022}), our two-stage approach can efficiently generate a doubly sparse solution, thereby constructing a highly accurate sparse model.
\end{remark}

%\subsection{Computational complexity}
%\label{subsec:complexity}

The following theorem gives an explicit upper bound for the computational complexity of the complete two-stage procedure in Algorithm~\ref{alg:3}.

\begin{theorem}\label{thm:5}
Suppose that the restricted CSQR refitting procedure with at most $q$
variables requires at most $L(q)$ Newton iterations. Then, the computational complexity of Algorithm~\ref{alg:3}
is
\[
\begin{aligned}
\mathrm{Cost}
=
O\Bigg\{
&
\sum_{s=1}^{s_{\max}}
N_1(s)
\Big[
L(sp_{\max})
\{ns^2p_{\max}^2+s^3p_{\max}^3\}
+np_{\max}p_n
+p_{\max}^2p_n
+J\log J
\Big]
\\
&
+
\sum_{r=1}^{r_{\max}}
N_2(r)
\Big[
L(r)(nr^2+r^3)
+n|\widehat{\mathcal S}|^\diamond+
|\widehat{\mathcal S}|^\diamond
\log|\widehat{\mathcal S}|^\diamond
\Big]
+
n(s_{\max}+r_{\max})
\Bigg\},
\end{aligned}
\]
where $N_1(s)$ is the number of outer active-set iterations of Algorithm~\ref{alg:1} and $N_2(r)$ is the number of outer active-set iterations of Algorithm~\ref{alg:2}.
%Here $|\widehat{\mathcal S}|^\diamond$ is the number of variables entering the second stage after the first-stage model size has been selected. In particular, since $|\widehat{\mathcal S}|^\diamond\le s_{\max}p_{\max}$, the second-stage term may be replaced by the corresponding deterministic upper bound obtained by substituting $s_{\max}p_{\max}$ for $|\widehat{\mathcal S}|^\diamond$.
\end{theorem}

\begin{remark}
Theorem~\ref{thm:5} characterizes the computational cost of the
proposed two-stage procedure. The first stage optimizes over active
group sets, while the second stage restricts variable selection to
the screened candidate set, whose dimension satisfies
$|\widehat{\mathcal S}|^\diamond\le s_{\max}p_{\max}\ll p_n$.
Thus, the procedure avoids exhaustive subset enumeration and
performs parameter updates on relatively low-dimensional active
sets. When the number of Newton iterations $L(q)$ and the
maximum active-set and group sizes remain bounded, the main
data-processing cost scales linearly with $n$ and $p_n$, apart
from sorting operations. These computational properties support
the applicability of the method to high-dimensional sparse problems.
\end{remark}

\section{Simulation studies}\label{sect.3A}

In this section, we conduct an extensive simulation study to evaluate the finite-sample performance of the proposed method. The simulation framework is designed to address two core aspects: (i) a sensitivity analysis of the smoothing kernel function and bandwidth; (ii) a comparative assessment against prevailing state-of-the-art methodologies, with comparisons made in terms of variable selection accuracy, parameter estimation precision, and computational efficiency.
All experiments were implemented in R software, and each configuration was replicated $200$ times to ensure statistical reliability.

\textbf{Example 1}. 
The samples $\{(\bm{\mathrm{x}}_i,\mathrm{y}_i)\}_{i=1}^n$
are generated from 
$
\mathrm{y}_i
=\alpha^*
+\sum_{j=1}^{J}
\bm{\mathrm{x}}_{i\mathcal G_j}^{\mathrm T}
\bm\beta_{\mathcal G_j}^*
+\epsilon_i
-F_{\epsilon_i\mid\bm{\mathrm{x}}_i}^{-1}(\tau), i=1,\ldots,n,$
where $F_{\epsilon_i|\bm{\mathrm{x}}_i}(\cdot)$ is the cumulative distribution function of $\epsilon_i$ given $\bm {\mathrm{x}}_i$. 
%Note that we subtract a shift $F_{\epsilon_i|\bm{\mathrm{x}}_i}^{-1}(\tau)$ for $\epsilon_i$ to ensure $\mathbb{P}\{\epsilon_i\leq 0|\bm {\mathrm{x}}_i\}=\tau$.
In this example, we consider a balanced group structure where each group contains $10$ variables (i.e., $\mathcal{G}_j=\{10\times(j-1)+1,\cdots,10\times j\}$ for $j=1,\cdots,J$). We randomly select a subset of active groups $\mathcal{S}^*$ from $\mathcal{C}=\{1,2,\cdots,J\}$ with $|\mathcal{S}^*|=s^*=5$ and $J=100$.
For the coefficients of the $j$th group $\bm\beta_{\mathcal{G}_j}^*=\big(\beta_{j'}^*,j' \in \mathcal{G}_j\big)^{\mathrm{T}}$, if $j\in \mathcal{F}^*=\mathcal{C}\backslash \mathcal{\mathcal{S}^*}$, we set $\bm\beta_{\mathcal{G}_j}^*=\bm 0$; and if $j\in \mathcal{S}^*$, we randomly take 5 elements of $\bm\beta^*_{\mathcal{G}_j}$ to be nonzeros (i.e., $\beta_{j'}^*$ is independently drawn either from ${\rm Uniform}(1,2)$ or ${\rm Uniform}(-2,-1)$) and the remaining elements in $\bm\beta^*_{\mathcal{G}_j}$ are all zeros. Thus, it is not hard to understand that there are five active groups, each containing 5 non-zero elements and 5 zero elements, implying that the total number of nonzero elements in $\bm\beta^*=\big(\bm\beta_{\mathcal{G}_j}^{*\mathrm{T}},j \in \mathcal{C}\big)^{\mathrm{T}}$ is 25.
The intercept term is taken as $\alpha^*=0.5$ here.
Following \cite{ZZZW23,LG2025}, we generate the $l$th element of the covariate vector $\bm{\mathrm{x}}_i$ by
$\mathrm{x}_{il} = \frac{\rho_1 \tilde{\mathrm{x}}_{ij} + z_{il}}{\sqrt{1 + \rho_1^2}}, \quad l = (j - 1)k + t,\quad 1 \leq i \leq n, \quad 1 \leq j \leq J, \quad 1 \leq t \leq k$ and $1 \leq l \leq p_n $,
where $\tilde{\bm{\mathrm{x}}}_i = (\tilde{\mathrm{x}}_{i1},\dots,\tilde{\mathrm{x}}_{iJ})^{\mathrm{T}} \overset{\text{i.i.d.}}{\sim} \mathcal{N}(\mathbf{0}, \bm\Sigma)$ with $\bm\Sigma = (\rho_2^{|j-j'|})_{1 \leq j, j' \leq J}$, $z_{i1},\ldots,z_{iJk}\overset{\text{i.i.d.}}{\sim}\mathcal{N}(0,1)$ and $k$ is group size. Then,
by a simple calculation, we have $\operatorname{Corr}(\mathrm{x}_{il}, \mathrm{x}_{il'}) =\dfrac{\rho_1^2}{1+\rho_1^2}$ if $l, l' \in \mathcal{G}_j$, and $\operatorname{Corr}(\mathrm{x}_{il}, \mathrm{x}_{il'}) =\dfrac{\rho_1^2 \rho_2^{|j-j'|}}{1+\rho_1^2}$ if $l \in \mathcal{G}_j,\; l' \in \mathcal{G}_{j'},\; j \neq j'$.
Thus, a large $\rho_1$ indicates strong correlation among variables within the same group, whereas $\rho_2$ solely impacts the correlation between different groups. Here we set $\rho_1=1$ and $\rho_2=0.5$, resulting in moderate within-group correlation and decaying group-wise correlation. Finally, the random error $\epsilon_i$ follows one of the following four distributions: (\emph{Case a}) standard normal distribution $\mathcal{N}(0,1)$;
(\emph{Case b}) a mixture distribution which is a mixture of $\mathcal{N}(0,1)$ and $\mathcal{N}(0,100)$ with weights $95\%$ and $5\%$ respectively; (\emph{Case c}) $t$-distribution with $3$ degrees of freedom; (\emph{Case d}) chi-squared distribution with $2$ degrees of freedom. \emph{Case b} and \emph{Case c} imply heavy-tailed errors and outliers, and \emph{Case d} stands for an asymmetric error.

To evaluate the performance across different approaches, we use the following criteria to
investigate accuracy of group selection, individual variable selection and parametric estimation:
\begin{itemize}
\item Group True-Positive Rate: ${\rm GTPR=GTP/\left( GTP+GFN \right)}$, where ${\rm GTP}=\left|\mathcal{S}^*\cap \hat{\mathcal{S}}\right|$ and ${\rm GFN}=\left|\mathcal{S}^*\cap \hat{\mathcal{F}}\right|$;

\item Group False-Positive Rate: ${\rm GFPR=GFP/\left( GFP+GTN \right)}$, where ${\rm GFP}=\left|\mathcal{F}^*\cap \hat{\mathcal{S}} \right|$ and ${\rm GTN}=\left|\mathcal{F}^*\cap \hat{\mathcal{F}}\right|$;

\item True-Positive Rate: ${\rm TPR=TP/\left( TP+FN \right)}$, where ${\rm TP}=\left|\mathcal{A}^*\cap \bar{\mathcal{A}}\right|$, ${\rm FN}=\left|\mathcal{A}^*\cap \bar{\mathcal{A}}^c\right|$ and $\bar{\mathcal{A}}^c = \Omega \backslash \bar{\mathcal{A}}$;

\item False-Positive Rate: ${\rm FPR=FP/\left( FP+TN \right)}$, where ${\rm FP}=\left|{\mathcal{A}^*}^c\cap \bar{\mathcal{A}} \right|$, ${\rm TN}=\left|{\mathcal{A}^*}^c\cap \bar{\mathcal{A}}^c\right|$ and ${\mathcal{A}^*}^c = \Omega \backslash \mathcal{A}^*$;

\item Estimation Error: ${{\rm  EE}=\left\|\left(\bar\alpha,\hat{\bm \beta}^{\mathrm{T}}\right)- \left(\alpha^*,\bm \beta^{*\mathrm{T}}\right)\right\|_2}$;

\item Out-of-sample Quantile Prediction Error: ${\rm{OQPE}}=\left| \mathcal{T} \right|^{-1}\sum\limits_{i \in \mathcal{T} }\rho_\tau\left(\mathrm{y}_i- \bar\alpha - \bm{\mathrm{x}}_{i}^{\mathrm{T}}\bar{\bm{\mathrm{\beta}}}\right)$, where $\mathcal{T}$ stands for a testing set with the size $\left| \mathcal{T} \right|$.
\end{itemize}

Recall that $\mathcal{S}^* = \mathrm{gsupp}(\bm{\mathrm{\beta}}^*) = \{ j\in \mathcal{C}: \|\bm{\mathrm{\beta}}^*_{\mathcal{G}_j}\|_2 \neq 0\}$, $\mathcal{F}^*=\mathcal{C}\backslash \mathcal{\mathcal{S}^*}$, $\mathcal{G}_{\mathcal{S}^*} = \bigcup_{j\in \mathcal{S}^*}\mathcal{G}_j$, $\mathcal{A}^* = \mathrm{supp}(\bm{\mathrm{\beta}}^*) = \{ k \in \mathcal{G}_{\mathcal{S}^*} : \beta^*_k \neq 0\}$ and $\mathcal{I}^*=\mathcal{G}_{\mathcal{S}^*}\backslash \mathcal{\mathcal{A}^*}$. Thus, {\rm GTPR} and {\rm GFPR} are employed to assess group selection accuracy, while {\rm TPR} and {\rm FPR} are adopted to evaluate individual variable selection accuracy. Notably, if $\bar{\mathcal{A}}=\mathcal{A}^*$ (i.e., {\rm TPR}=1 and {\rm FPR}=0), it implies all important groups are not omitted in the first stage (i.e. {\rm GTPR} has to be one).
The remaining two criteria serve to quantify estimation accuracy.
For each simulation, we randomly allocate 20$\%$ of the total samples to the testing set, with the remaining 80$\%$ serving as the training set.
We estimate unknown parameters, choose significant variables and determine tuning parameters by using samples from the training set, and use samples from the testing set to calculate {\rm{OQPE}}.
This allows us to evaluate the out-of-sample performance of various approaches.

It is evident that the outputs of {\sf SGSA-HDQR} (Algorithm \ref{alg:3}) may depend on selection of the kernel function $K(\cdot)$ and smoothing parameter $h$. We conduct a sensitivity analysis for several commonly used kernel functions (e.g., Gaussian, Uniform, Logistic and Epanechnikov) with different bandwidths $h=0.2,0.4,0.6,0.8,1$. We consider $\tau=0.25, 0.5, 0.75$ and $n=300$. Specifically, we consider the following four kernel functions:
\begin{itemize}

\item Gaussian kernel: $K(u)=\frac{1}{\sqrt{2\pi}}\exp\left(-u^2/2\right)$, the resulting smoothed loss takes the form $\ell_h(u)=h\phi(u/h) - u\Phi(-u/h) + \tau u$, where $\phi(\cdot)$ and $\Phi(\cdot)$ denote the probability density function and cumulative distribution function of the standard normal distribution, respectively.

\item Uniform kernel: $K(u)=(1/2)\mathbb{I}(|u|\leq 1)$, the resulting smoothed loss is $\ell_h(u)=\frac{1}{4h}(u^2+h^2)\mathbb{I}(|u|\leq h)+\frac{|u|}{2}\mathbb{I}(|u|> h) + (\tau-\frac{1}{2})u$.

\item Logistic kernel: $K(u)=\exp(-u)/(1+\exp(-u))^2$, the resulting smoothed loss is $\ell_h(u)=\tau u+h\log(1+\exp(-u/h))$.

\item Epanechnikov kernel: $K(u)=(3/4)(1-u^2)\mathbb{I}(|u|\leq 1)$, the resulting smoothed loss is $\ell_h(u)=\frac{h}{2}(\frac{3u^2}{4h^2}-\frac{u^4}{8h^4}+\frac{3}{8})\mathbb{I}(|u|\leq h) + \frac{|u|}{2}\mathbb{I}(|u|>h) + (\tau-\frac{1}{2})u$.
\end{itemize}

\tabcolsep=1.5pt
\begin{table}[htbp]
\centering
\caption{\small Simulation results for Example 1. The smallest EE and OQPE are flagged with the symbol $\sharp$.
The symbol ``-" indicates that the result of a method is not available.}
\label{table3}
\small
\begin{tabular}{l ccccccc c cccccc}
\hline
\noalign{\smallskip}
\multirow{3}{*}{Method}& GTPR & GFPR & TPR & FPR & EE & OQPE && GTPR & GFPR & TPR & FPR & EE & OQPE \\
\cline{2-14}
& \multicolumn{13}{c}{$\tau=0.25$}\\
& \multicolumn{6}{c}{\emph{Case a}} && \multicolumn{6}{c}{\emph{Case b}} \\

{\sf TS-Lasso}    & 1.000 & 0.618 & 0.978 & 0.408 & 5.281 & 1.297 && 1.000 & 0.462 & 0.981 & 0.347 & 5.376 & 1.445 \\
{\sf TS-SCAD}     & 1.000 & 0.027 & 1.000 & 0.020 & 0.726 & 0.373 && 1.000 & 0.030 & 1.000 & 0.024 & 0.869 & 0.563 \\
{\sf TS-MCP}      & 1.000 & 0.026 & 1.000 & 0.022 & 0.747 & 0.373 && 1.000 & 0.029 & 1.000 & 0.024 & 0.886 & 0.567 \\
{\sf GSDRFA}      & 1.000 & 0.000 & - & - & 0.780 & 0.369 && 0.996 & 0.000 & - & - & 0.939 & 0.558 \\
{\sf SSDRFA}      & - & - & 1.000 & 0.001 & 0.540 & 0.348 && - & - & 0.916 & 0.001 & 1.387 & 0.666 \\
{\sf SGSA-HDQR}   & 1.000 & 0.000 & 1.000 & 0.000 & 0.533$^\sharp$ & 0.345$^\sharp$ && 0.996 & 0.000 & 0.996 & 0.000 & 0.620$^\sharp$ & 0.525$^\sharp$ \\

& \multicolumn{6}{c}{\emph{Case c}} && \multicolumn{6}{c}{\emph{Case d}} \\
{\sf TS-Lasso}    & 1.000 & 0.541 & 0.976 & 0.379 & 5.370 & 1.387 && 1.000 & 0.439 & 0.982 & 0.332 & 5.262 & 1.415 \\
{\sf TS-SCAD}     & 1.000 & 0.038 & 1.000 & 0.027 & 1.059 & 0.546 && 1.000 & 0.024 & 1.000 & 0.020 & 0.728 & 0.490 \\
{\sf TS-MCP}      & 1.000 & 0.034 & 1.000 & 0.027 & 1.068 & 0.543 && 1.000 & 0.025 & 1.000 & 0.022 & 0.737 & 0.492 \\
{\sf GSDRFA}      & 1.000 & 0.000 & - & - & 1.094 & 0.525 && 1.000 & 0.000 & - & - & 0.847 & 0.496 \\
{\sf SSDRFA}      & - & - & 0.980 & 0.002 & 0.995 & 0.539 && - & - & 0.962 & 0.001 & 0.918 & 0.537 \\
{\sf SGSA-HDQR}   & 1.000 & 0.000 & 1.000 & 0.000 & 0.708$^\sharp$ & 0.487$^\sharp$ && 1.000 & 0.000 & 1.000 & 0.000 & 0.498$^\sharp$ & 0.457$^\sharp$ \\

\cline{2-14}
& \multicolumn{13}{c}{$\tau=0.5$}\\
& \multicolumn{6}{c}{\emph{Case a}} && \multicolumn{6}{c}{\emph{Case b}} \\
{\sf TS-Lasso}    & 1.000 & 0.608 & 0.997 & 0.489 & 4.615 & 1.446 && 1.000 & 0.494 & 0.998 & 0.413 & 4.700 & 1.600 \\
{\sf TS-SCAD}     & 1.000 & 0.034 & 1.000 & 0.020 & 0.664 & 0.457 && 1.000 & 0.032 & 1.000 & 0.021 & 0.717 & 0.642 \\
{\sf TS-MCP}      & 1.000 & 0.027 & 1.000 & 0.016 & 0.645 & 0.452 && 1.000 & 0.027 & 1.000 & 0.019 & 0.724 & 0.642 \\
{\sf GSDRFA}      & 1.000 & 0.000 & - & - & 0.729 & 0.456 && 1.000 & 0.000 & - & - & 0.801 & 0.641 \\
{\sf SSDRFA}      & - & - & 1.000 & 0.000 & 0.492 & 0.431 && - & - & 1.000 & 0.000 & 0.510 & 0.606 \\
{\sf SGSA-HDQR}   & 1.000 & 0.000 & 1.000 & 0.000 & 0.480$^\sharp$ & 0.428$^\sharp$ && 1.000 & 0.000 & 1.000 & 0.000 & 0.504$^\sharp$ & 0.605$^\sharp$ \\

& \multicolumn{6}{c}{\emph{Case c}} && \multicolumn{6}{c}{\emph{Case d}} \\
{\sf TS-Lasso}    & 1.000 & 0.556 & 0.997 & 0.461 & 4.809 & 1.583 && 1.000 & 0.522 & 0.996 & 0.437 & 4.922 & 1.678 \\
{\sf TS-SCAD}     & 1.000 & 0.036 & 1.000 & 0.022 & 0.818 & 0.627 && 1.000 & 0.041 & 1.000 & 0.026 & 1.124 & 0.782 \\
{\sf TS-MCP}      & 1.000 & 0.034 & 1.000 & 0.022 & 0.819 & 0.629 && 1.000 & 0.036 & 1.000 & 0.025 & 1.131 & 0.784 \\
{\sf GSDRFA}      & 1.000 & 0.000 & - & - & 0.918 & 0.627 && 1.000 & 0.000 & - & - & 1.171 & 0.768 \\
{\sf SSDRFA}      & - & - & 0.992 & 0.000 & 0.659 & 0.603 && - & - & 0.969 & 0.001 & 1.080 & 0.784 \\
{\sf SGSA-HDQR}   & 1.000 & 0.000 & 1.000 & 0.000 & 0.572$^\sharp$ & 0.583$^\sharp$ && 1.000 & 0.000 & 1.000 & 0.000 & 0.773$^\sharp$ & 0.723$^\sharp$ \\

\cline{2-14}
& \multicolumn{13}{c}{$\tau=0.75$}\\
& \multicolumn{6}{c}{\emph{Case a}} && \multicolumn{6}{c}{\emph{Case b}} \\
{\sf TS-Lasso}    & 1.000 & 0.620 & 0.978 & 0.406 & 5.278 & 1.304 && 1.000 & 0.470 & 0.981 & 0.351 & 5.434 & 1.442 \\
{\sf TS-SCAD}     & 1.000 & 0.031 & 1.000 & 0.021 & 0.739 & 0.379 && 1.000 & 0.032 & 1.000 & 0.025 & 0.876 & 0.555 \\
{\sf TS-MCP}      & 1.000 & 0.024 & 1.000 & 0.019 & 0.744 & 0.380 && 1.000 & 0.031 & 1.000 & 0.025 & 0.904 & 0.558 \\
{\sf GSDRFA}      & 1.000 & 0.000 & - & - & 0.795 & 0.374 && 1.000 & 0.000 & - & - & 0.921 & 0.544 \\
{\sf SSDRFA}      & - & - & 1.000 & 0.001 & 0.571 & 0.356 && - & - & 0.924 & 0.001 & 1.337 & 0.645 \\
{\sf SGSA-HDQR}   & 1.000 & 0.000 & 1.000 & 0.000 & 0.541$^\sharp$ & 0.349$^\sharp$ && 1.000 & 0.000 & 1.000 & 0.000 & 0.592$^\sharp$ & 0.512$^\sharp$ \\

& \multicolumn{6}{c}{\emph{Case c}} && \multicolumn{6}{c}{\emph{Case d}} \\
{\sf TS-Lasso}    & 1.000 & 0.556 & 0.975 & 0.385 & 5.482 & 1.422 && 1.000 & 0.570 & 0.957 & 0.377 & 5.768 & 1.545 \\
{\sf TS-SCAD}     & 1.000 & 0.041 & 1.000 & 0.030 & 1.065 & 0.562 && 0.999 & 0.074 & 0.998 & 0.052 & 2.224 & 0.915 \\
{\sf TS-MCP}      & 1.000 & 0.039 & 1.000 & 0.031 & 1.105 & 0.566 && 0.999 & 0.070 & 0.999 & 0.052 & 2.284 & 0.931 \\
{\sf GSDRFA}      & 1.000 & 0.000 & - & - & 1.066 & 0.539 && 0.953 & 0.000 & - & - & 2.235 & 0.879 \\
{\sf SSDRFA}      & - & - & 0.958 & 0.001 & 1.157 & 0.587 && - & - & 0.602 & 0.002 & 4.604 & 1.267 \\
{\sf SGSA-HDQR}   & 1.000 & 0.000 & 1.000 & 0.000 & 0.707$^\sharp$ & 0.502$^\sharp$ && 0.953 & 0.000 & 0.941 & 0.001 & 1.816$^\sharp$ & 0.840$^\sharp$ \\
\noalign{\smallskip}\hline
\end{tabular}
\end{table}

The numerical results from Figure S1 of the Supplementary Material indicate that group-wise and within-group sparsity for {\sf SGSA-HDQR} (Algorithm \ref{alg:3}) is insensitive to the choice of kernel function.
Figure S1 also shows the logistic kernel yields larger estimation errors than other three kernels. Moreover, we find that {\sf SGSA-HDQR} is insensitive to the choice of $h$ because the numerical results exhibit little variation across different $h$ values.
This aligns with the findings of \cite{YZZ2023,HPTZ2023}. To effectively trade off between bias and variance, we suggest adopting the Gaussian kernel in practice, and take the bandwidth as $h=\frac{1}{2}\left(\frac{|\mathcal{S}^0|^\diamond\log p_n}{n}\right)^{1/5}$ with $|\mathcal{S}^0|^\diamond=\sum_{j \in\mathcal{S}^0}p_j$ according to the suggestion of \cite{YZZ2023}, where $\mathcal{S}^0$ is the initial set of active groups in Algorithm \ref{alg:3}.

Table S1 of the Supplementary Material lists the difference between {\sf SGSA-HDQR} and oracle estimator in terms of EE and OQPE.
The oracle estimator is based on the prior information that we know which are the relevant variables in advance, and it is of no practical utility and can only be used as a benchmark in simulations.
We see that the difference between {\sf SGSA-HDQR} and oracle estimator exhibits a consistent decreasing trend and approaches to zero as the sample size increases. This demonstrates that estimation and prediction accuracy of {\sf SGSA-HDQR} is gradually close to that of the oracle estimator, thereby validating the robustness and estimation efficiency of {\sf SGSA-HDQR} in scenarios involving outliers, heavy-tailed and asymmetric errors.

To make full comparison, we compare the proposed {\sf SGSA-HDQR} with the two-stage penalized quantile regression with popular $\ell_1$ and nonconvex penalties \citep{PW2015}.
Specifically, in the first stage, we achieve group-wise sparsity via penalized quantile regression using group Lasso, SCAD, or MCP penalties. Subsequently, in the second stage, we further identify important variables within these active groups by applying penalized quantile regression with element-wise Lasso, SCAD, or MCP penalty. We refer these two-stage methods as {\sf TS-Lasso}, {\sf TS-SCAD} and {\sf TS-MCP}, which can be easily implemented by the R package ``rqPen". Moreover, to assess the necessity of the proposed double sparsity framework, two simplified variants of {\sf SGSA-HDQR} are also examined here.
The first variant, denoted as {\sf GSDRFA}, is implemented using Algorithm \ref{alg:1} with the GBIC criterion to select the active groups, but omits the subsequent within-group selection step, thereby enforcing only group-wise sparsity.
The second variant, denoted as {\sf SSDRFA}, disregards the grouping structure entirely and directly applies Algorithm \ref{alg:2} with the BIC criterion to select the sparsity level, performing individual variable selection without leveraging group information. These two baselines serve to highlight the distinct contributions of group-wise and within-group sparsity, as well as the advantage of incorporating both simultaneously in the full procedure.

Table \ref{table3} presents simulation results of all compared methods for Example 1.
From the perspective of variable selection, {\sf TS-Lasso} consistently selects an excessive number of irrelevant variables, leading to inflated GFPR and FPR in almost all scenarios. Although {\sf TS-SCAD} and {\sf TS-MCP} partially mitigate over-selection, they still fail to fully control false discoveries, particularly for non-normal error distributions. In contrast, {\sf SGSA-HDQR} maintains near-perfect selection accuracy across most experimental settings, simultaneously achieving high true positive rates and negligible false positives, which suggests a stable identification of both relevant groups and important variables within groups.
In terms of estimation and prediction performance, {\sf SGSA-HDQR} attains the smallest EE and OQPE in most settings. Notably, its advantage becomes more pronounced for heavy-tailed and asymmetric error distributions, where traditional two-stage methods tend to incorporate numerous insignificant variables, leading to substantial degradation in estimation accuracy.

Furthermore, a comparison of {\sf SGSA-HDQR} with its two simplified variants highlights the limitations of a single sparse structure. {\sf GSDRFA} only achieves group-wise sparsity, retains redundant variables within groups, directly leading to larger EE and OQPE.
On the other hand, {\sf SSDRFA} disregards the grouping structure and performs individual variable selection in isolation, exhibiting a significant decline in estimation accuracy, especially for non-normal error distributions.
The proposed {\sf SGSA-HDQR} achieves both group-wise and within-group sparsity simultaneously, enabling it to accurately eliminate redundant information in high-dimensional complex data and ultimately achieve superior estimation accuracy, model parsimony, and prediction robustness.

\tabcolsep=1.5pt
\begin{table}[htbp]
\centering
\caption{\small Simulation results for Example 2. The smallest EE and OQPE are flagged with the symbol $\sharp$.
The symbol ``-" indicates that the result of a method is not available.}
\label{table3C}
\small
\begin{tabular}{l ccccccc c cccccc}
\hline
\noalign{\smallskip}
\multirow{3}{*}{Method}& GTPR & GFPR & TPR & FPR & EE & OQPE & & GTPR & GFPR & TPR & FPR & EE & OQPE \\
\cline{2-14}
& \multicolumn{13}{c}{$\tau=0.25$}\\
& \multicolumn{6}{c}{\emph{Case a}} && \multicolumn{6}{c}{\emph{Case b}} \\

{\sf TS-Lasso}    &0.810 &0.308 &0.780 &0.275 &9.417 &1.918 & &0.732 &0.246 &0.697 &0.230 &9.620 &2.025 \\
{\sf TS-SCAD}     &0.740 &0.083 &0.734 &0.106 &5.842 &1.347 & &0.540 &0.047 &0.518 &0.058 &7.691 &1.746 \\
{\sf TS-MCP}      &0.659 &0.066 &0.653 &0.085 &6.261 &1.408 & &0.453 &0.042 &0.423 &0.048 &8.444 &1.895 \\
{\sf GSDRFA}      &0.996 &0.000 &- &- &1.142 &0.433 & &0.896 &0.000 &- &- &2.939 &0.907\\
{\sf SSDRFA}      &- &- &0.009 &0.000 &10.617 &2.000 & &- &- &0.009 &0.000 &10.650 &2.053\\
{\sf SGSA-HDQR}   &0.996 &0.000 &0.996 &0.000 &0.795$^\sharp$ &0.380$^\sharp$ & &0.896 &0.000 &0.893 &0.003 &2.544$^\sharp$ &0.846$^\sharp$\\

& \multicolumn{6}{c}{\emph{Case c}} && \multicolumn{6}{c}{\emph{Case d}} \\
{\sf TS-Lasso}    &0.753 &0.257 &0.720 &0.241 &9.566 &1.992 & &0.723 &0.257 &0.695 &0.240 &9.501 &2.016 \\
{\sf TS-SCAD}     &0.618 &0.060 &0.605 &0.075 &7.056 &1.639 & &0.554 &0.053  &0.530 &0.066 &7.525 &1.728 \\
{\sf TS-MCP}      &0.491 &0.045 &0.467 &0.053 &8.068 &1.764 & &0.468 &0.044 &0.446 &0.052 &8.092 &1.805  \\
{\sf GSDRFA}      &0.988 &0.000 &- &- &1.721 &0.646 & &0.970 &0.000 &- &- &1.80 &0.661\\
{\sf SSDRFA}      &- &- &0.008 &0.000 &10.605 &2.047 & &- &- &0.008 &0.000 &10.421 &2.055\\
{\sf SGSA-HDQR}   &0.988 &0.000 &0.988 &0.001 &1.203$^\sharp$ &0.566$^\sharp$ & &0.970 &0.000 &0.969 &0.001 &1.265$^\sharp$ &0.569$^\sharp$ \\

\cline{2-14}
& \multicolumn{13}{c}{$\tau=0.5$}\\
& \multicolumn{6}{c}{\emph{Case a}} && \multicolumn{6}{c}{\emph{Case b}} \\
{\sf TS-Lasso}    &1.000 &0.900 &0.776 &0.631 &9.457 &2.414 & &0.975 &0.759 &0.793 &0.560 &9.395 &2.484 \\
{\sf TS-SCAD}     &0.999 &0.133 &0.999 &0.164 &3.287 &1.080 & &0.954 &0.124 &0.952 &0.152 &4.261 &1.474\\
{\sf TS-MCP}      &0.997 &0.132 &0.997 &0.163 &3.214 &1.059 & &0.930 &0.108 &0.926 &0.139 &4.305 &1.477 \\
{\sf GSDRFA}      &1.000 &0.000 &- &- &1.012 &0.503 & &0.992 &0.000 &- &- &1.281 &0.753\\
{\sf SSDRFA}      &- &- &0.009 &0.000 &10.004 &2.485 & &- &- &0.009 &0.000 &10.005 &2.563\\
{\sf SGSA-HDQR}   &1.000 &0.000 &1.000 &0.000 &0.685$^\sharp$ &0.455$^\sharp$ & &0.992 &0.000 &0.992 &0.000 &0.873$^\sharp$ &0.682$^\sharp$ \\

& \multicolumn{6}{c}{\emph{Case c}} && \multicolumn{6}{c}{\emph{Case d}} \\
{\sf TS-Lasso}    &0.997 &0.831 &0.818 &0.623 &9.442 &2.466 & &0.986 &0.804 &0.765 &0.572 &9.496 &2.161 \\
{\sf TS-SCAD}     &0.974 &0.149 &0.973 &0.176 &4.436 &1.946 & &0.950 &0.135 &0.945 &0.161 &4.930 &1.673
\\
{\sf TS-MCP}      &0.927 &0.123 &0.922 &0.149 &4.488 &1.489 & &0.943 &0.129 &0.935 &0.152 &4.913 &1.641 \\
{\sf GSDRFA}      &1.000 &0.000 &- &- &1.326 &0.707 & &0.888 &0.000 &- &- &3.238 &1.202\\
{\sf SSDRFA}      &- &- &0.009 &0.000 &10.006 &2.539 & &- &- &0.009 &0.000 &10.018  &2.580\\
{\sf SGSA-HDQR}   &1.000 &0.000 &1.000 &0.000 &0.851$^\sharp$ &0.628$^\sharp$ & &0.888 &0.000 &0.886 &0.003 &2.790$^\sharp$ &1.124$^\sharp$ \\

\cline{2-14}
& \multicolumn{13}{c}{$\tau=0.75$}\\
& \multicolumn{6}{c}{\emph{Case a}} && \multicolumn{6}{c}{\emph{Case b}} \\
{\sf TS-Lasso}    &0.852 &0.314 &0.823 &0.287 &9.387 &1.904 & &0.717 &0.234 &0.692 &0.223 &9.618 &2.059 \\
{\sf TS-SCAD}     &0.794 &0.085 &0.786 &0.110 &5.352 &1.265 & &0.594 &0.055 &0.578 &0.068 &7.425 &1.781 \\
{\sf TS-MCP}      &0.709 &0.073 &0.698 &0.093 &5.978 &1.359 & &0.453 &0.040 &0.438 &0.049 &8.237 &1.853 \\
{\sf GSDRFA}      &1.000 &0.000 &- &- &1.089 &0.422 & &0.904 &0.000 &- &- &2.813 &0.882\\
{\sf SSDRFA}      &- &- &0.008 &0.000 &10.598 &1.953 & &- &- &0.009 &0.000 &10.628 &2.097\\
{\sf SGSA-HDQR}   &1.000 &0.000 &1.000 &0.000 &0.731$^\sharp$ &0.369$^\sharp$ & &0.904 &0.000 &0.901 &0.003 &2.424$^\sharp$ &0.824$^\sharp$ \\

& \multicolumn{6}{c}{\emph{Case c}} && \multicolumn{6}{c}{\emph{Case d}} \\
{\sf TS-Lasso}    &0.768 &0.256 &0.744 &0.245 &9.469 &1.967 & &0.694  &0.228 &0.661 &0.219 &9.622 &2.016 \\
{\sf TS-SCAD}     &0.667 &0.066 &0.657 &0.085 &6.714 &1.572 & &0.400 &0.039 &0.366 &0.041 &8.899 &1.918 \\
{\sf TS-MCP}      &0.529 &0.051 &0.511 &0.064 &7.820 &1.747 & &0.331 &0.030 &0.304 &0.034 &9.147 &1.945 \\
{\sf GSDRFA}      &0.975 &0.000 &- &- &1.960 &0.694 & &0.648 &0.000 &- &- &6.975 &1.618 \\
{\sf SSDRFA}      &- &- &0.008 &0.000 &10.597 &2.026 & &- &- &0.008 &0.000 &10.627 &2.033\\
{\sf SGSA-HDQR}   &0.975 &0.000 &0.974 &0.001 &1.439$^\sharp$ &0.612$^\sharp$ & &0.648 &0.000 &0.635 &0.012 &6.845$^\sharp$ &1.512$^\sharp$ \\
\noalign{\smallskip}\hline
\end{tabular}
\end{table}

\textbf{Example 2}. To evaluate the adaptability of the proposed method to such structural heterogeneity, we now consider a setting with varying group sizes. Specifically, we assume that the size of the $j$th group is independently drawn from a discrete uniform distribution ${\rm U}\{15,\ldots,20\}$ for $j=1,\cdots,J$, implying $p_j\overset{\text{i.i.d.}}{\sim} {\rm U}\{15,\ldots,20\}$ and $p_n=\sum_{j=1}^Jp_j$. The covariates $\bm{\mathrm{x}}_i$ are generated from a $p_n$-dimensional multivariate normal distribution $\mathcal{N}(0,\bm\Xi)$, where the $(j,l)$th element of $\bm\Xi$ equals  $0.5^{|j-l|}$ for $1\leq j,l \leq p_n$. The true coefficient vector $\bm\beta_{\mathcal{G}_j}^*=\big(\beta_{j'}^*,j' \in \mathcal{G}_j\big)^{\mathrm{T}}$ for the $j$th group is constructed as follows. We take the $j'$th element of $\bm\beta_{\mathcal{G}_j}^*$ as $\beta_{j'}^*=(-1)^{j'}$
for $j\in \mathcal{S}^*$ and $j'=1,\cdots,10$, while $\beta_{j'}^*=0$ for others. Set $J=200$ and $\mathcal{S}^*=\{1,2,\cdots,10\}$. Therefore, it is easy to see that the first 10 groups are important and the first 10 variables within these groups are nonzeros.
The intercept term is taken as $\alpha^*=-0.5$.
In this example, we consider a higher-dimensional setting with more active groups and more variables within these active groups.
Other settings are the same as in Example 1.

%In Figure \ref{figureA1}, we report the running time (in seconds) of two types of sparse methods for Example 2: group-wise sparsity approaches (including Algorithm \ref{alg:1}, group Lasso and group SCAD) and within-group sparsity approaches (including Algorithm \ref{alg:2}, Lasso and SCAD), with the variation of the parameter $s$ (or $\lambda$).
%Note that SCAD and MCP are both nonconvex penalties, and hence their running time is similar. So, we do not report the running time of group MCP and MCP here for saving space.
%We can see from the left panel of Figure \ref{figureA1} that the running time of traditional group Lasso and group SCAD increase almost sublinearly as the $\lambda$ decreases, while the running time of Algorithm \ref{alg:1} almost varies little and is much less than the other two approaches. Additionally, the right panel of Figure \ref{figureA1} indicates that the various within-group sparsity methods show negligible differences in their running time. Hence, when evaluated against other two-stage competing methods, the proposed {\sf SGSA-HDQR} (Algorithm \ref{alg:3}) exhibits a more pronounced superiority in terms of computational efficiency.

Table \ref{table3C} reports that our proposal has satisfactory numerical performances and outperforms other competitors.
For instance, when $\tau=0.25$, {\sf SGSA-HDQR} attains both group-wise and within-group sparsity precisely for \emph{Case c}, as evidenced by its GTPR and TPR of 98.8$\%$ alongside near-zero GFPR and FPR. In contrast, none of the competing methods can reliably achieve these dual sparsity properties, owing to their low GTPR and TPR values coupled with non-negligible GFPR and FPR.
Overall, the results clearly demonstrate that {\sf SGSA-HDQR} has satisfactory performance in sparse-group selection.

\section{Analysis of breast cancer gene expression data}\label{sect.4A}

In this section, we illustrate our proposal by a breast cancer gene expression dataset, which can be accessed at \url{http://myweb.uiowa.edu/pbreheny/data/bcTCGA.html}.
The dataset comprises expression profiles of 17,814 mRNA microarrays across 536 patients, with all expression values recorded on a logarithmic scale. Prior to analysis, 491 genes with missing values were excluded.
Previous studies have shown that the ``BRCA1" is the gene associated with the increased risk of early-onset breast cancer \citep{SHJY19}. Hence we take it as the response variable. Given that the ``BRCA1" is known to interact with numerous other genes, it is of great interest to identify genes whose expression levels are associated with the ``BRCA1" expression.
Several researchers (e.g., \cite{SHJY19,MY2024,TY24}) have studied
the same data set to analyze the mean of ``BRCA1".
Unlike previous studies that focus on mean regression,
we aim to identify genes most strongly associated with ``BRCA1" and investigate how the selected genes affect ``BRCA1" expression at different quantiles.
Figure S2 of the Supplementary Material presents the histogram and Q-Q plot of ``BRCA1" expression, showing an approximately normal distribution with minor deviations in the tails and the presence of outliers.
The presence of outliers in ``BRCA1" expression provides a rationale for employing quantile regression in this study.

Although the dataset contains a large number of genes, only a small subset may be associated with ``BRCA1" expression. Following \cite{TY24,MY2024}, we employ the quantile correlation method of \cite{LLT15} to screen the 2,000 genes most strongly associated with ``BRCA1" expression. These genes are then grouped according to the C1 positional gene sets in the Molecular Signatures Database (MSigDB), which classify genes based on their chromosomal locations, following \cite{SFHT2013,LAL2019}. Table S2 of the Supplementary Material summarizes the resulting group structures at three quantile levels. Since the screened genes may differ across quantile levels, the corresponding group structures also vary. Moreover, the substantial variation in group sizes indicates that the dataset exhibits an unbalanced group structure.

%In this empirical study, we employ the Gaussian kernel for the proposed method and adopt the same bandwidth selection scheme as in Section \ref{sect.3A}.

\begin{table}[htbp]
\centering
\footnotesize
\caption{\small The selection frequency of the top 15 genes and average of OQPE over 100 replications, with standard deviation in parentheses. The underlined genes represent the common genes identified by the four methods.}
\label{table5}
\begin{tabular}{cclclclcl}
\toprule
\multirow{2}{*}{No.} & \multicolumn{2}{c}{\sf TS-Lasso} & \multicolumn{2}{c}{\sf TS-SCAD} & \multicolumn{2}{c}{\sf TS-MCP} & \multicolumn{2}{c}{\sf SGSA-HDQR} \\
\cmidrule(lr){2-3} \cmidrule(lr){4-5} \cmidrule(lr){6-7} \cmidrule(lr){8-9}
 & Gene & Freq. & Gene & Freq. & Gene & Freq. & Gene & Freq. \\
\hline
&\multicolumn{8}{c}{$\tau = 0.25$} \\
1 & \underline{NBR2}  & 100  & \underline{NBR2} & 100 & \underline{NBR2}  & 100 & \underline{NBR2} & 100 \\
2 & \underline{TOP2A} & 100 & \underline{TOP2A} & 100 & \underline{CENPK} & 100 & \underline{CENPK} & 82 \\
3 & \underline{CENPK} & 100 & \underline{CENPK} & 100 & \underline{TOP2A} & 99 & \underline{TOP2A} & 81 \\
4 & \underline{VPS25} & 99  & \underline{VPS25} & 98 & \underline{VPS25} & 89  & \underline{SPRY2} & 77 \\
5 & \underline{CDC6} & 96 & \underline{CDC6} & 93 & NPEPPS & 75 & USP9Y & 63 \\
6 & PSME3 & 90 & NPEPPS & 88 & PSME3 & 65 & \underline{CDC6} & 42 \\
7 & NPEPPS & 84 & CCDC43 & 86 & CCDC43 & 53 & BECN1 & 38 \\
8 & CCDC43 & 80 & PSME3 & 82 & \underline{LSM12} & 48 & \underline{PLEKHH3} & 27 \\
9 & \underline{PLEKHH3}  & 80 & \underline{LSM12} & 71 & \underline{CDC6} & 45 & \underline{LSM12} & 26 \\
10 & NMT1 & 68 & \underline{PLEKHH3} & 71 & \underline{PLEKHH3} & 45 & \underline{SPAG5} & 24 \\
11 & \underline{SPRY2} & 63 & MLX & 60 & \underline{HMMR} & 37 & \underline{HMMR} & 24 \\
12 & \underline{LSM12} & 61 & \underline{HMMR} & 48 & \underline{SPAG5} & 34 & GINS1 & 22 \\
13 & \underline{HMMR} & 54 & \underline{SPRY2}  & 48 & NSF & 30 & \underline{VPS25} & 20 \\
14 & NSF & 47 & NSF & 46  & \underline{SPRY2} & 29 & DEPDC1B & 7 \\
15 & \underline{SPAG5} & 35 & \underline{SPAG5} & 35 & NMT1 & 27 & RUNDC1 & 5 \\
OQPE  & \multicolumn{2}{c}{0.166(0.017)} & \multicolumn{2}{c}{0.161(0.017)} & \multicolumn{2}{c}{0.156(0.016)} & \multicolumn{2}{c}{0.158(0.012)} \\
\hline
&\multicolumn{8}{c}{$\tau = 0.5$} \\
1 & \underline{NBR2} & 100  & \underline{NBR2} & 100 & \underline{NBR2}  & 100 & \underline{NBR2} & 100 \\
2 & CENPK & 100  & CENPK & 100  & CENPK   & 100 & \underline{HMMR} & 100 \\
3 & \underline{HMMR} & 100 & \underline{HMMR} & 99 & \underline{TOP2A}  & 99 &\underline{CCDC56} & 88 \\
4 & \underline{CCDC56} & 100  &\underline{CCDC56} & 99  & \underline{LSM12} & 98 & CDC6 & 78 \\
5 & \underline{TOP2A} & 100  & \underline{TOP2A} & 99 & \underline{HMMR}   & 94 &\underline{TOP2A} & 70 \\
6 & \underline{LSM12} & 100 & \underline{LSM12} & 99  & \underline{CCDC56} & 94  & \underline{SPRY2} & 67 \\
7 & PSME3  & 100  & PSME3 & 99  & KPNB1  & 93  &\underline{LSM12} & 60 \\
8 & GCN5L2 & 99   &GCN5L2 & 98  & PSME3  & 90  & DTL  & 48 \\
9 & \underline{EFTUD2} & 96 & KPNB1 & 94  & GCN5L2 & 72 &\underline{VPS25} & 47 \\
10 & KPNB1 & 95 & \underline{NMT1} & 89 & ZWINT  & 65  & ASPM & 36 \\
11 & CDC6  & 94 & \underline{VPS25} & 89  & \underline{NMT1} & 64 &KIF14  & 23 \\
12 &\underline{VPS25} & 92 & \underline{EFTUD2} & 85 & \underline{VPS25}  & 63 & \underline{NMT1}   & 18 \\
13 & \underline{NMT1} & 85 & CDC6  & 83  & \underline{EFTUD2} & 57  & PLEKHH3  & 14 \\
14 & \underline{SPRY2} & 81 & \underline{SPRY2} & 73  & \underline{SPRY2} & 43  & \underline{EFTUD2} & 13 \\
15 & ZWINT & 49   & ZWINT & 65  & PLEKHH3 & 37 &  AARSD1   & 13 \\
OQPE  & \multicolumn{2}{c}{0.174(0.017)} & \multicolumn{2}{c}{0.174(0.018)} & \multicolumn{2}{c}{0.172(0.018)} & \multicolumn{2}{c}{0.166(0.016)} \\
\hline
&\multicolumn{8}{c}{$\tau = 0.75$} \\
1 & ZWINT & 100  & \underline{NBR2} & 100 & \underline{TOP2A}  & 100 & \underline{NBR2} & 100 \\
2 & PSME3 & 99 & PSME3 & 100 & PSME3 & 99  & HMMR & 97 \\
3 & \underline{TOP2A} & 99  & \underline{TOP2A} & 100 & \underline{NBR2} & 98 & \underline{CDC6} & 94 \\
4 & \underline{NBR2} & 99 & ZWINT & 99 & \underline{VPS25} & 91 & ASPM & 55 \\
5 & EFTUD2 & 98 & EFTUD2 & 99  & ZWINT & 90 & NBR1 & 54 \\
6 & \underline{VPS25} & 98  & \underline{VPS25} & 98  & EFTUD2 & 84 & PLEKHH3 & 34 \\
7 & \underline{MCM6} & 96  & \underline{MCM6} & 95 & \underline{MCM6}  & 82 & \underline{VPS25} & 32 \\
8 & TUBG1 & 91 & TUBG1 & 88  & \underline{LSM12} & 51 & \underline{LSM12} & 31 \\
9 & \underline{CDC6} & 68 & \underline{LSM12} & 70  & CENPK & 48 & SPRY2 & 28 \\
10 & \underline{LSM12} & 67 & \underline{CDC6} & 66  & TUBG1 & 29 & \underline{MCM6} & 23 \\
11 & HMGB2  & 55 & HMGB2 & 51  & \underline{CDC6} & 27  & \underline{TOP2A} & 20 \\
12 & DBF4B & 48 & CENPK & 48 & \underline{OIP5} & 24 & RUNDC1 & 19 \\
13 & WDR76 & 43 & WDR76 & 42 & WDR76 & 20 & FIGNL1 & 14 \\
14 & \underline{OIP5} & 41 & DBF4B & 41 & PROS1 & 18  & RPL27 & 13 \\
15 & CENPK & 37 & \underline{OIP5} & 40  & FANCI & 17 & \underline{OIP5} & 11 \\
OQPE  & \multicolumn{2}{c}{0.144(0.014)} & \multicolumn{2}{c}{0.144(0.012)} & \multicolumn{2}{c}{0.141(0.013)} & \multicolumn{2}{c}{0.140(0.012)} \\
\bottomrule
\end{tabular}
\end{table}

\begin{figure}[htbp]
    \centering
    \renewcommand{\arraystretch}{1.0}
    \setlength{\tabcolsep}{0.5pt}
    \begin{tabular}{ccc}
    \subfloat{\includegraphics[width=0.32\textwidth]{"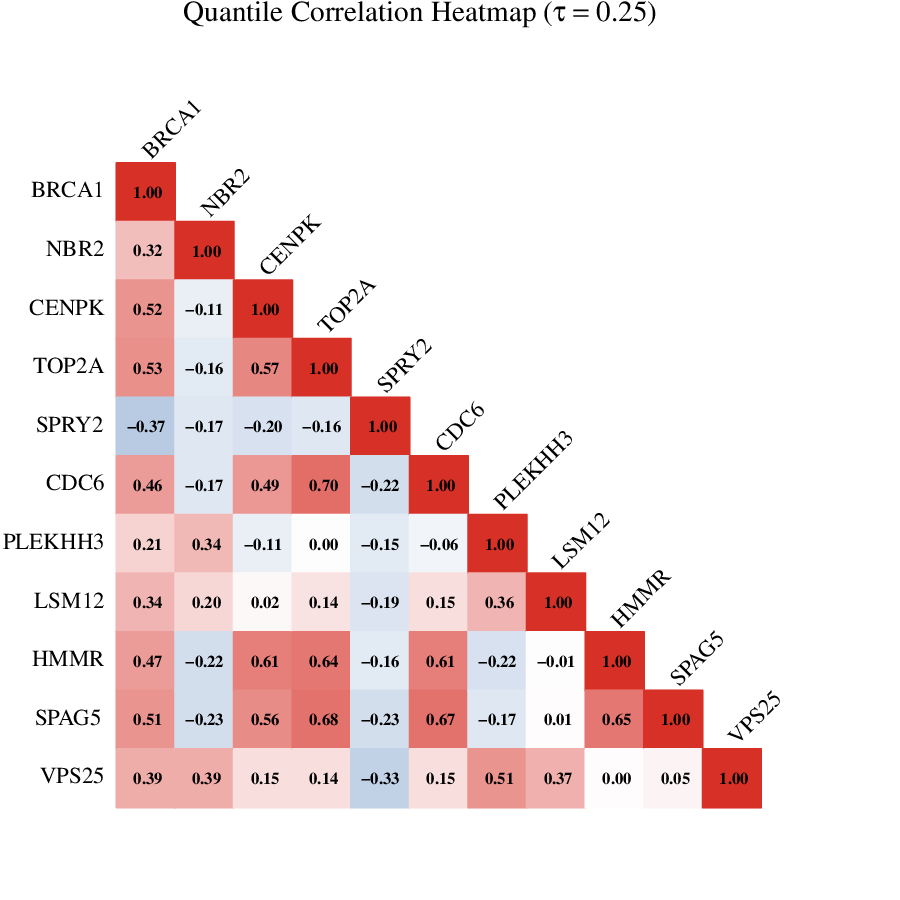"}}&
    \subfloat{\includegraphics[width=0.32\textwidth]{"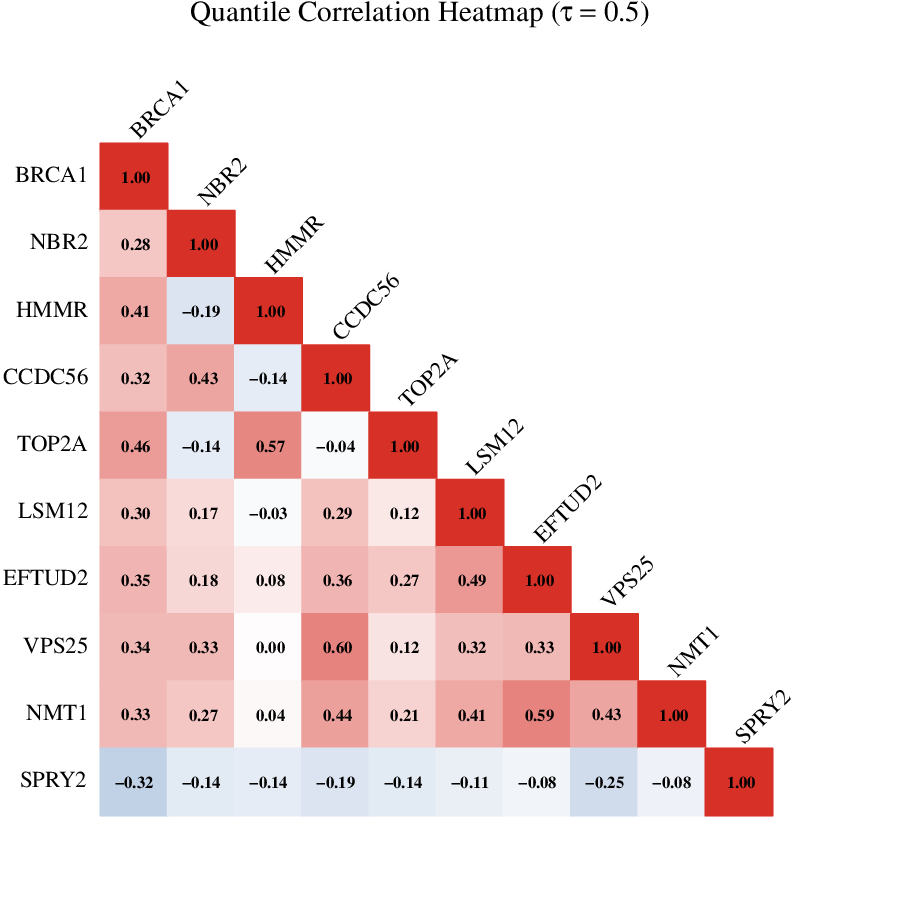"}}&
    \subfloat{\includegraphics[width=0.32\textwidth]{"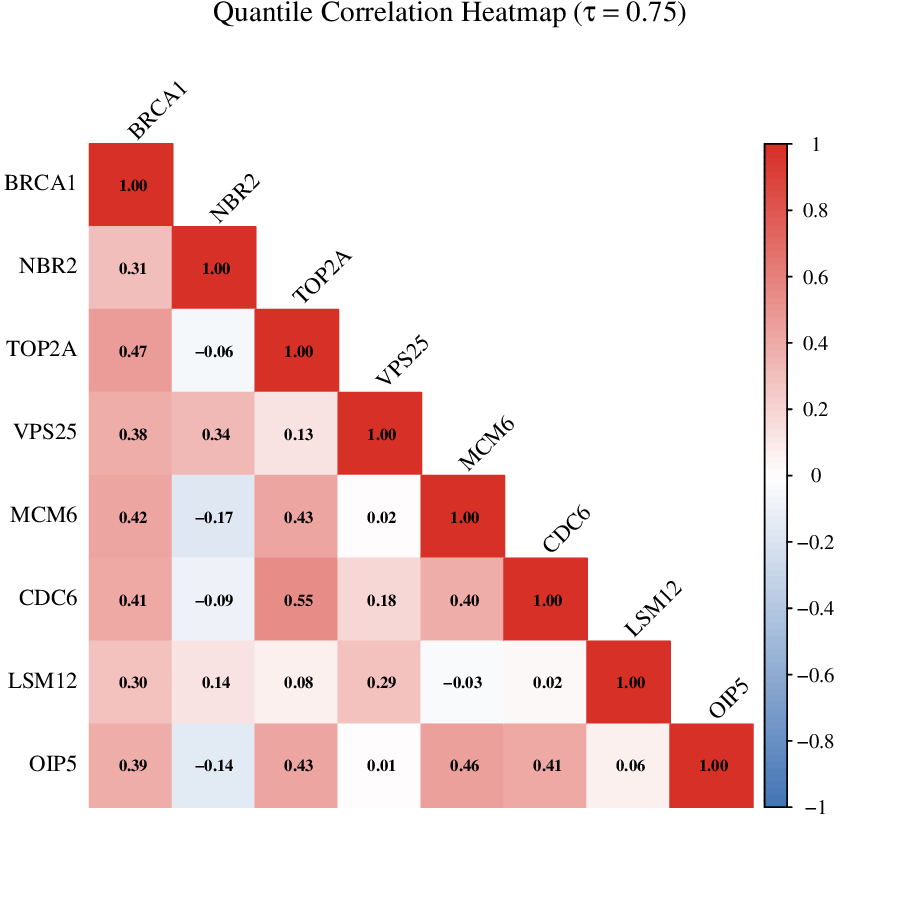"}}\\
   \end{tabular}
   \caption{\small The quantile correlation heatmaps between ``BRCA1" and common genes identified by the four methods for $\tau=0.25$, 0.5 and 0.75.}
  \label{figure4}
\end{figure}

To evaluate performance, we randomly split the 536 patients into a training set of 436 samples and a test set of the remaining 100 samples. We identify important genes and fit the model on the training set, and evaluate the out-of-sample quantile prediction error (OQPE) defined in Section \ref{sect.3A} on the test set to assess predictive performance. This process is repeated 100 times. Table \ref{table5} lists the top 15 most frequently selected genes across all methods, together with the average OQPE over 100 replications.
As shown in Table \ref{table5}, our proposed {\sf SGSA-HDQR} exhibits highly competitive out-of-sample predictive performance. Specifically, it attains the lowest OQPE values (0.166 and 0.140) among competing approaches at $\tau = 0.5$ and $\tau = 0.75$. Although its OQPE (0.158) is marginally higher than that of {\sf TS-MCP} (0.156) at $\tau = 0.25$, the performance remains quantitatively comparable and statistically competitive. Importantly, {\sf SGSA-HDQR} yields the smallest standard deviations among all methods, which clearly demonstrates its enhanced stability and robustness over repeated replications.

Despite different variable selection approaches may choose different important genes for a given quantile level, there are some overlapped genes by various methods. Table \ref{table5} also reports several overlapped genes that are chosen by the four methods. For example, the overlapped genes selected by the four methods for $\tau=0.75$ are ``NBR2", ``CDC6", ``VPS25", ``LSM12", ``MCM6", ``TOP2A" and ``OIP5", indicating these genes may have significant influence on the BRCA1 expression at $\tau=0.75$.
Figure \ref{figure4} depicts the quantile correlation between the ``BRCA1" and overlapped genes by the four methods at $\tau = 0.25, 0.5$ and 0.75.
We observe that some genes maintain relatively strong correlations across all quantile levels, whereas others exhibit pronounced associations only at specific quantiles.
Meanwhile, the strength of correlation for individual genes varies noticeably across quantile levels. Such heterogeneous correlation patterns further underscore the complex regulatory mechanisms underlying ``BRCA1" expression.

\begin{figure}[htbp]
    \centering
    \renewcommand{\arraystretch}{1.0}
    \setlength{\tabcolsep}{0.5pt}
    \begin{tabular}{ccc}
    \subfloat{\includegraphics[width=0.32\textwidth]{"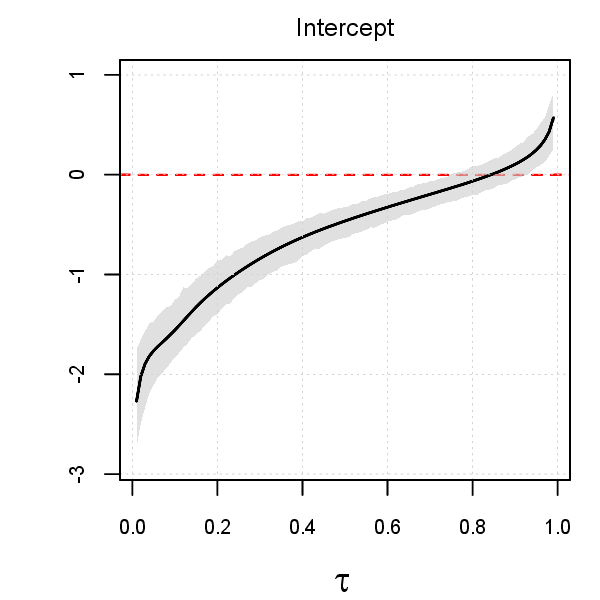"}}&
    \subfloat{\includegraphics[width=0.32\textwidth]{"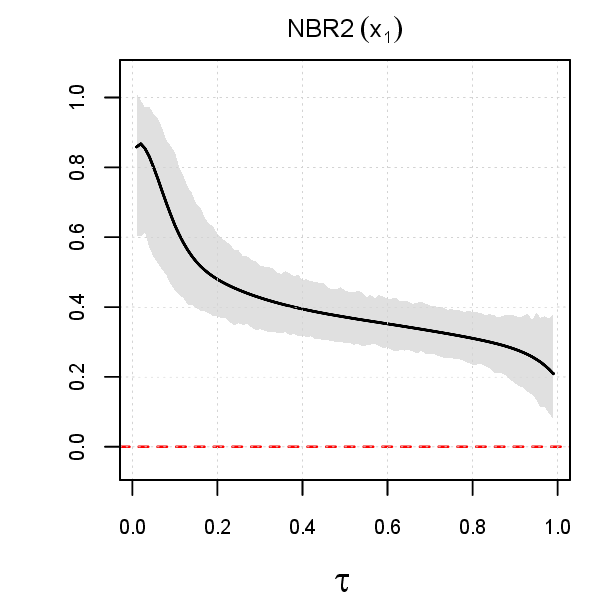"}} &
    \subfloat{\includegraphics[width=0.32\textwidth]{"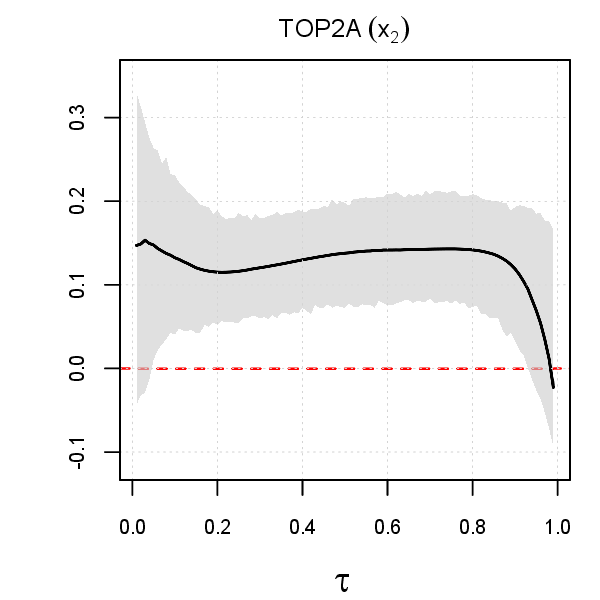"}} \\
    \subfloat{\includegraphics[width=0.32\textwidth]{"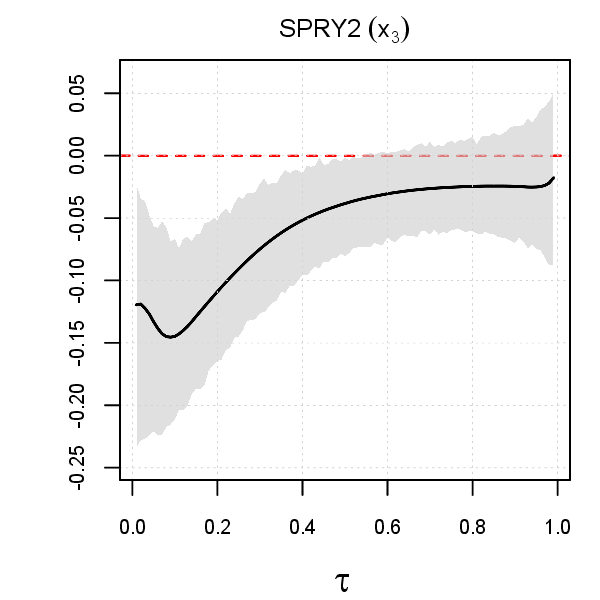"}} &
    \subfloat{\includegraphics[width=0.32\textwidth]{"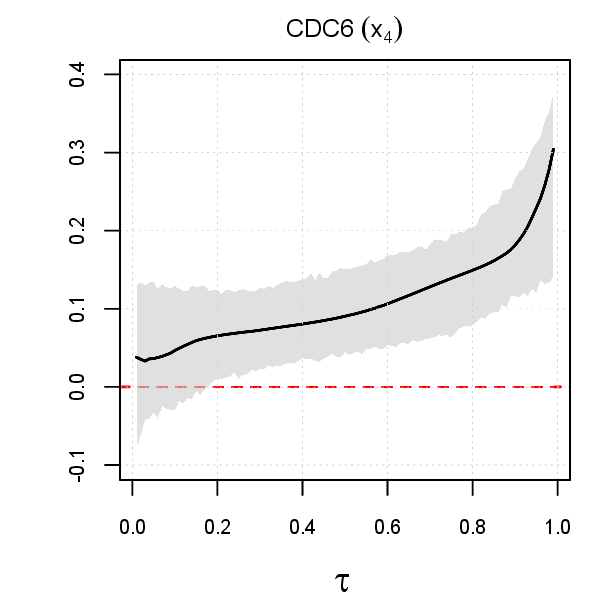"}} &
    \subfloat{\includegraphics[width=0.32\textwidth]{"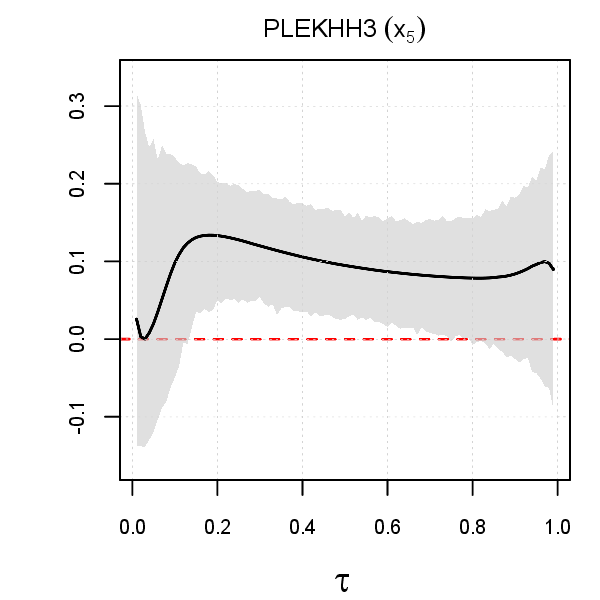"}} \\
    \subfloat{\includegraphics[width=0.32\textwidth]{"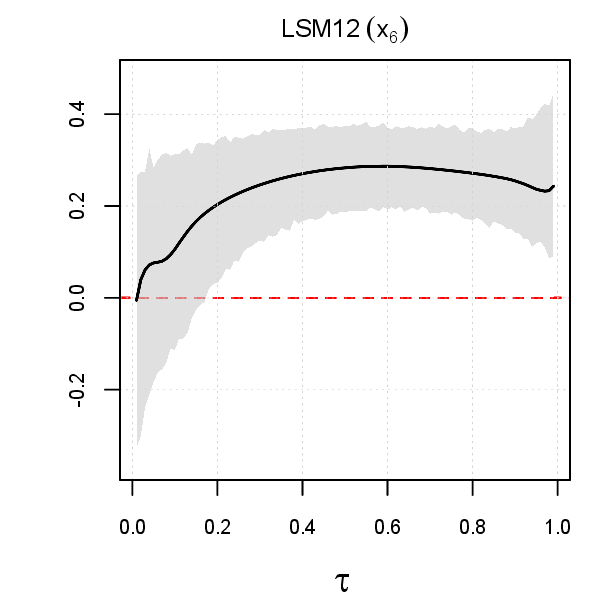"}} &
    \subfloat{\includegraphics[width=0.32\textwidth]{"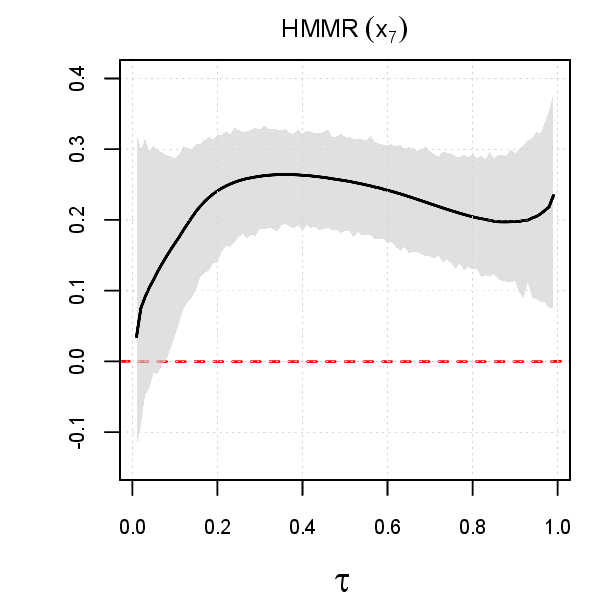"}} &
    \subfloat{\includegraphics[width=0.32\textwidth]{"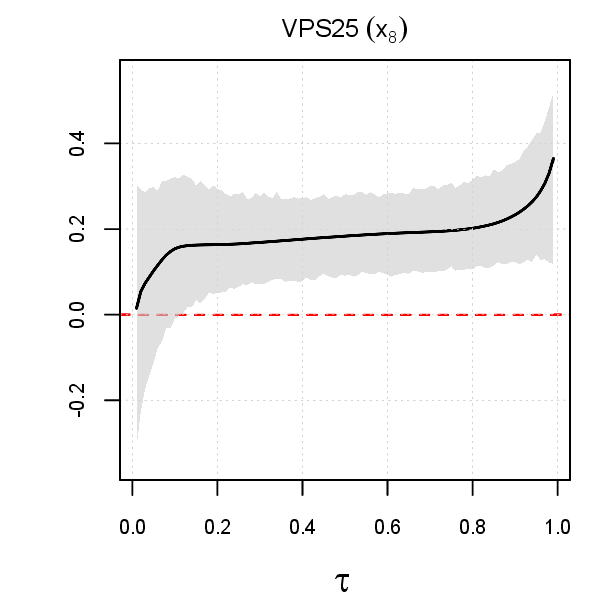"}} \\

   \end{tabular}
   \caption{\small The estimated regression coefficients (black solid lines) and their 95\% confidence intervals (shaded areas) for the eight overlapped genes in analysis of breast cancer gene expression data.}
  \label{figure5*}
\end{figure}

Furthermore, it should be noted that the selection of important genes may also vary with quantile levels. Nevertheless, ``NBR2", ``TOP2A", ``SPRY2", ``CDC6", ``PLEKHH3", ``LSM12", ``HMMR" and ``VPS25" are identified as important common genes by {\sf SGSA-HDQR} for the three quantile levels. More interestingly, ``NBR2" is selected by {\sf SGSA-HDQR} with the highest frequency across the three quantile levels. Previous studies such as \cite{BP2019} and \cite{SHJY19} have shown that ``NBR2" is unquestionably associated with ``BRCA1" because they are adjacent to one another on chromosome 17 and share a promoter. Thus our proposal {\sf SGSA-HDQR} provides strong biological plausibility for selection of important genes. To further examine the magnitude and statistical significance of the overlapped genes (including ``NBR2" (${\mathrm{x}}_1$), ``TOP2A" (${\mathrm{x}}_2$), ``SPRY2" (${\mathrm{x}}_3$), ``CDC6" (${\mathrm{x}}_4$), ``PLEKHH3" (${\mathrm{x}}_5$), ``LSM12" (${\mathrm{x}}_6$), ``HMMR" (${\mathrm{x}}_7$) and ``VPS25" (${\mathrm{x}}_8$)), we consider the following parsimonious model
\begin{align*}
Q_{\tau}\left(\mathrm{ y}_i|\bm {\mathrm{x}}_i\right)=\alpha_{\tau}+
\sum_{j =1}^{8 } {\mathrm{x}}_{ij} \beta_{j,\tau},~~ i=1,\cdots,536,
\end{align*}
where $\bm {\mathrm{x}}_i=\left(\mathrm{x}_{i1},\cdots,\mathrm{x}_{i8} \right)^{\mathrm{T}}$. Here, we consider a set of quantiles with $\tau \in \{0.01,0.02,\cdots,0.99\}$.
We can estimate $\alpha_{\tau},\beta_{1,\tau},\cdots,\beta_{8,\tau}$ by minimizing the convoluted smoothed quantile loss
$\widetilde{\mathcal{L}}_{n,h}(\alpha_{\tau},\bm{\mathrm{\beta}}_{\tau})$ given in Section 2.2, which can be easily implemented by the function ``conquer" in the R package ``conquer" \citep{HPTZ2023}.
Figure \ref{figure5*} displays the estimated regression coefficients and their 95\% confidence intervals.
It is readily observed that the confidence intervals for all genes are relatively narrow and exclude zero at most of quantile levels, suggesting that the eight genes exert statistically significant regulatory effects on ``BRCA1" expression. Moreover, it is not hard to understand from Figure \ref{figure5*} that
``SPRY2" shows a negative association with ``BRCA1" expression, with confidence intervals strictly below zero at most of quantile levels. Conversely, the remaining seven genes exhibit positive regulatory effects, as their respective confidence intervals lie entirely above zero at the majority of quantile ranges. Overall, results from the real data analysis validate that {\sf SGSA-HDQR} is highly effective for simultaneous achievement of both group-wise
and within-group sparsity, clearly outperforming existing competitors.

\section{Concluding Remarks}\label{sect.5A}

In this paper, we develop a robust two-stage sparse-group selection method for high-dimensional quantile regression that simultaneously achieves group-wise and within-group sparsity. Algorithm~\ref{alg:1} combines the convoluted smoothed quantile loss with the $\ell_{2,0}$ penalty for group selection, after which Algorithm~\ref{alg:2} imposes an $\ell_0$ penalty within the retained groups; the two stages, together with the GBIC and BIC selectors, are integrated into the adaptive {\sf SGSA-HDQR} procedure (Algorithm~\ref{alg:3}). Under mild regularity conditions, both stages enjoy sure screening and exact recovery guarantees, and the information criteria are selection consistent.
Numerical experiments and the breast cancer gene-expression analysis show that {\sf SGSA-HDQR} outperforms the competing {\sf TS-Lasso}, {\sf TS-SCAD}, and {\sf TS-MCP} methods in selection accuracy, estimation accuracy, and computational efficiency, identifies biologically meaningful genes, and produces reliable quantile predictions. Overall, it provides a useful addition to the toolkit for high-dimensional quantile regression.

Several directions merit further investigation. First, this paper assumes a non-overlapping group structure, whereas in applications such as gene pathway analysis, genes often participate in multiple pathways simultaneously; extending the framework to overlapping groups is therefore of substantial practical interest. Second, a distributed computing implementation could alleviate the memory and computational bottlenecks for ultra-large-scale datasets. Third, building on the smoothed quantile loss, it would be of interest to develop post-selection inferential tools, such as confidence intervals and hypothesis tests for the quantile coefficients in the doubly sparse model.

\section*{Acknowledgments}
This paper is supported by National Natural Science Foundation of China (Grant No. 12571290), Fundamental Research Funds for the Central Universities (Grant No. SWU-KU24002).

%\clearpage
%\bibliography{ref}
%\bibliographystyle{apalike}

\end{document}